\documentclass[%
reprint,
superscriptaddress,
twocolumn,
showpacs,
amsmath,amssymb,
aps,
floatfix,
showkeys,
]{revtex4-2}

\usepackage{graphicx}
\usepackage{dcolumn}
\usepackage{bm}
\usepackage{amsmath}
\usepackage{amssymb}
\usepackage{multirow}
\usepackage{color}
\usepackage[T1]{fontenc}
\usepackage[colorlinks=true,linkcolor=blue,citecolor=blue]{hyperref}
\usepackage{booktabs}
\usepackage{tabularx}
\usepackage[english]{babel}
\usepackage[utf8]{inputenc}
\usepackage{enumerate}
\usepackage[cmyk,dvipsnames]{xcolor}

\usepackage[normalem]{ulem}

\date{ \today}

\begin{document}

\title{Reduction of energy cost of magnetization switching in a biaxial nanoparticle by use of internal dynamics}

\author{Mohammad H.A. Badarneh}
\email[Corresponding author: ]{mha5@hi.is}
\affiliation{Science Institute of the University of Iceland, 107 Reykjav\'ik, Iceland}

\author{Grzegorz J. Kwiatkowski}
\affiliation{Science Institute of the University of Iceland, 107 Reykjav\'ik, Iceland}

\author{Pavel F. Bessarab}
\affiliation{Science Institute of the University of Iceland, 107 Reykjav\'ik, Iceland}
\affiliation{Department of Physics and Electrical Engineering, Linnaeus University, SE-39231 Kalmar, Sweden}

\begin{abstract}
A solution to energy-efficient magnetization switching in a nanoparticle with biaxial anisotropy is presented. Optimal control paths minimizing the energy cost of magnetization reversal are calculated numerically as functions of the switching time and materials properties, and used to derive energy-efficient switching pulses of external magnetic field. Hard-axis anisotropy reduces the minimum energy cost of magnetization switching due to the internal torque in the desired switching direction. Analytical estimates quantifying this effect are obtained based on the perturbation theory. The optimal switching time providing a tradeoff between fast switching and energy efficiency is obtained. The energy cost of switching and the energy barrier between the stable states can be controlled independently in a biaxial nanomagnet. This provides a solution to the dilemma between energy-efficient writability and good thermal stability of magnetic memory elements. 
\end{abstract}

\maketitle

\section{\label{intro}Introduction}

Identification of energy limits for the control of magnetization is an important fundamental problem of condensed matter physics. It is also a prerequisite for the development of energy-efficient technologies based on magnetic materials. An important application is magnetic memory where writing of data is realized via selective magnetization reversals in nanoelements.  Magnetization reversal can be achieved by various means, including optical pulses \cite{vomir2005real,hohlfeld2001fast,le2012demonstration}, spin-polarized electric current \cite{myers1999current,katine2000current}, external magnetic~\cite{sun2005fast,acremann2000imaging,hiebert1997direct,xiao2006minimal} and electric field \cite{yang2017ultrafast}, microwave-assisted reversal switching \cite{thirion2003switching,sun2006magnetization,cai2013reversal}, stress \cite{bandyopadhyay2021magnetic}, temperature gradient \cite{Pushp2015,Michel_2017}, etc. The challenge is to minimize the energy cost of the control stimulus generation.

In conventional bit recording, magnetization reversal in a memory element is achieved by applying a static external magnetic field in an opposite direction to the initial magnetization. This results in a relatively slow reversal process governed by damping as long as the magnitude of the external field exceeds the coercive field~\cite{bertotti2003comparison, mallinson2000damped}. The coercive field and, thereby, the energy cost of switching can be reduced by decreasing the magnetic anisotropy, but this may lead to unwanted reversals induced by thermal fluctuations due to decrease in the energy barrier separating the stable states. One solution to this dilemma between good thermal stability and energy-efficient writability of magnetic elements for memory applications is use of exchange spring magnets~\cite{suess2009exchange}, where the energy barrier and the coercive field can be tuned independently.

Decrease in the switching time and/or the switching field can also be achieved via realization of special reversal protocols such as precessional magnetization switching~\cite{back1998magnetization}. Precessional switching is typically induced by applying a magnetic field pulse transverse to the initial magnetization, but the pulse duration must be chosen accurately so as to avoid back switching~\cite{bertotti2003critical}. Additionally, precessional switching is prone to instabilities due to the magnetization ringing effect~\cite{bauer2000suppression} unless the switching pulse is properly shaped~\cite{gerrits2002ultrafast,bauer2000suppression, schumacher2003phase}. In microwave-assisted reversals, the switching field can also be reduced thanks to resonant energy pumping~\cite{thirion2003switching,rivkin2006magnetization,cai2013reversal, sun2006magnetization}. 

Clearly, the possibility to achieve the reversal by several different methods implies the existence of an optimal protocol, but its definite identification is a challenging problem. Barros \textit{et al.} employed the optimal control theory (OCT)~\cite{pontryagin2018mathematical} to establish a formal approach to the magnetization switching optimization~\cite{barros2011optimal,barros2013microwave}. Within the approach, the optimal switching pulse is found as a result of a direct minimization of the switching cost functional under the constraint defined by a system-specific magnetization dynamics. In our previous article, we revisited the OCT due to Barros \textit{et al.} using unconstrained minimization, which helped us find a complete analytical solution to the energy-efficient reversal of a nanomagnet with uniaxial anisotropy~\cite{kwiatkowski2021optimal}. 

\begin{figure}[!t]
\centering
\includegraphics[width=\columnwidth]{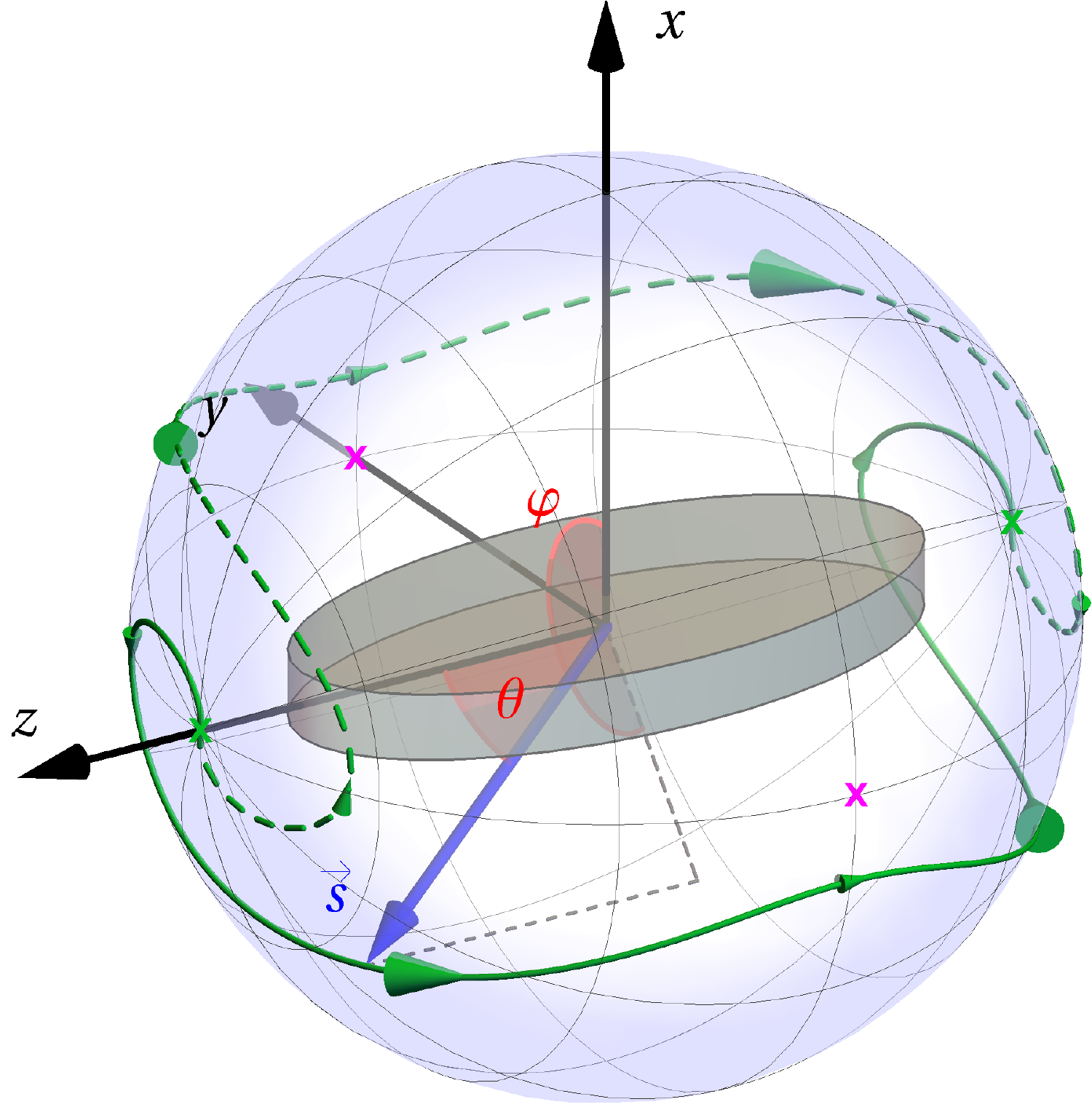}
\caption{\label{system_picture}Optimal switching of a flat elongated nanomagnet representing a biaxial anisotropy system. The direction of the normalized magnetic moment $\vec{s}$ is shown with the blue arrow. Orientations of $\vec{s}$ that correspond to the minima and the saddle points on the energy surface are marked with the green and magenta crosses, respectively. The calculated optimal control paths between the energy minima are shown with the solid and the dashed green lines. The damping factor $\alpha$ is 0.1, the switching time $T$ is $8\tau_0$, and the hard-axis anisotropy constant is twice as large as the easy-axis anisotropy constant. The green arrows along the reversal paths show the velocity of the system at $t=T/6$, $t=T/3$, $t=T/2$, $t=2T/3$, and $t=5T/6$, with the size of the arrowheads being proportional to the magnitude of the velocity. The contours of constant azimuthal angle $\varphi$ (meridians) and polar angle $\theta$ (parallels) are shown with thin black lines.}
\end{figure}

We also reported decrease in the switching cost for systems with biaxial anisotropy, the result of the internal torque produced by the hard axis~\cite{kwiatkowski2021optimal}. That the internal torque can assist magnetization reversal was recognized earlier for several systems, for example for Co films~\cite{back1999minimum} and
Co nanoclusters~\cite{etz2012accelerating}. The aim of the present study is to explore this effect quantitatively. We focus on nanomagnets with biaxial anisotropy, which can arise due to the demagnetizing field~\cite{osborn1945demagnetizing}. This scenario is realized in flat elongated nanoelements; see Fig.~\ref{system_picture}. Such systems are used, e.g., as single bits in in-plane memory~\cite{chun2012scaling}, or as elements of artificial spin ice arrays~\cite{nisoli2013colloquium,wysin2012magnetic}. 

We investigate by means of the OCT to what extent the energy cost of magnetization switching can be minimized by pulse shaping and how this depends on the parameters of the biaxial system and the switching time. Thanks to the internal torque generated by the hard-axis anisotropy, the energy cost can be reduced below the free-macrospin level. Based on the perturbation theory, we show some analytical estimates of the energy cost reduction. We show that in a biaxial system the energy barrier separating the stable states and energy cost of switching between them can be tuned independently, which provides a solution to the magnetic recording dilemma.

The article is organized as follows. Sec.~\ref{methods} provides a theoretical framework for energy-efficient control of magnetization by means of external magnetic field: In Sec.~\ref{methods:Optimal control theory}, the OCT for magnetic systems is presented and the corresponding Euler-Lagrange equation for the optimal control path (OCP), a dynamical trajectory minimizing the energy cost of magnetization switching, is derived; In Sec.~\ref{methods:Numerical method for finding OCP}, the numerical method for finding OCPs and corresponding energy-efficient control pulses via direct minimization of the cost functional is presented; In Sec.~\ref{methods:Perturbation theory}, a method for finding an approximate solution for the minimum energy cost is worked out based on the perturbation theory. The application of the methodology to a biaxial anisotropy system is presented in Sec.~\ref{results}. Conclusions and discussion are presented in Sec. \ref{conclusion}.

\section{Methodology} 
\label{methods}

\subsection{Optimal control theory} 
\label{methods:Optimal control theory}
We define the cost of the magnetization switching as the amount of energy used to generate the control pulse that produces the desired change in the magnetic structure of the system. Assuming the control to be an external magnetic field generated by an electric circuit, the energy cost is mostly defined by Joule heating due to the resistance of the circuit. This is proportional to the square of the electric current integrated over the switching time. Taking into account the linear relationship between the current magnitude and the strength of the generated field, the cost functional can be written as~\cite{barros2011optimal,kwiatkowski2021optimal,badarneh2020mechanisms}
 \begin{equation} \label{eq:one}
    \Phi = \int_0^T 
    |\vec{B}(t)|^2dt,
\end{equation}
where $T$ is the switching time and $\vec{B}(t)$ is the generated external magnetic field at time $t$. The aim of the OCT is to identify the optimal pulse $\vec{B}_m(t)$ that brings the system to the desired final state such that $\Phi$ is minimized. Whenever thermal fluctuations are negligible, the system dynamics can be described by the Landau-Lifshitz-Gilbert (LLG) equation~\cite{berkov2007magnetization}:
\begin{equation} \label{LLG}
\left(1+\alpha^2\right) \dot{\vec{s}} = -\gamma \vec{s} \times \bigl(\vec{b}+\vec{B}\bigr)  - \alpha \gamma  \vec{s} \times \left[\vec{s} \times \bigl(\vec{b}+\vec{B}\bigr) \right],
\end{equation}
where $\vec{s}$ is the normalized magnetic moment vector, $\gamma$ is the gyromagnetic ratio, $\alpha$ is the damping factor, and $\vec{b}$ is the internal magnetic field defined by the magnetic configuration through the following equation:
\begin{equation}
 \vec{b} = \vec{b}(\vec{s})= -\frac{1}{\mu}\frac{\partial E}{\partial  \vec{s} } 
\end{equation}
with $\mu$ being the magnetic moment length and $E$ the internal energy of the system.

Both $\vec{B}(t)$ and $\vec{s}(t)$ can be treated as independent variables, and $\Phi$ minimized subject to the constraint defined by Eq.~(\ref{LLG})~\cite{barros2011optimal,barros2013microwave}. Alternatively, the optimal pulse $\vec{B}_m(t)$ can be calculated via unconstrained minimization of $\Phi$. For this, Eq.~(\ref{LLG}) is used to express the external magnetic field in terms of the dynamical trajectory and the internal magnetic field~\cite{kwiatkowski2021optimal}: 
\begin{eqnarray} \label{eq:three}
    \vec{B}(\vec{s},\dot{\vec{s}}) = \frac{\alpha}{\gamma}\dot{\vec{s}}+\frac{1}{\gamma}\left[\vec{s}\times \dot{\vec{s}}\right]-\vec{b}^{\perp},
\end{eqnarray}
with $
\vec{b}^\perp = \vec{b} - \vec{s}\bigl(\vec{b}\cdot \vec{s}\bigr)$ being the transverse component of the internal field, and the result substituted into~Eq.~(\ref{eq:one}). Subsequently, the energy cost  $\Phi$ becomes a functional of the switching trajectory $\vec{s}(t)$:
\begin{equation} \label{eq:cost_funct}
\Phi=\Phi[\vec{s}(t)]=\int_0^T A(\vec{s},\dot{\vec{s}}) dt, 
\end{equation}
where $A(\vec{s},\dot{\vec{s}})$ is given by
\begin{equation} \label{eq:func}
    A(\vec{s},\dot{\vec{s}})=\frac{\alpha^2+1}{\gamma^2}|\dot{\vec{s}}|^2 - \frac{2 \alpha}{\gamma} \dot{\vec{s}} \cdot \vec{b}^\perp -\frac{2}{\gamma}\left(\vec{s}\times \dot{\vec{s}}\right)\cdot \vec{b}^\perp + |\vec{b}^\perp|^2. 
\end{equation}
The optimal reversal mechanism can be found by minimizing $\Phi$ with respect to path connecting the initial and the final state in the configuration space. Corresponding OCP $\vec{s}_m(t)$ can be identified by solving the Euler-Lagrange equation:
\begin{equation}\label{eq:EL}
    \begin{split}
    \left[\left(\vec{s}\cdot\vec{b}\right) \hat{I}-\frac{1}{\mu}\hat{H}\right]\left[\frac{1}{\gamma}\vec{s}\times\dot{\vec{s}}-\vec{b}^{\perp}\right] &  \\  +\frac{1}{\mu}\left[\vec{s}\cdot\hat{H}\left(\frac{1}{\gamma}\vec{s}\times\dot{\vec{s}}-\vec{b}^{\perp}\right)\right]\vec{s} &  \\-\frac{1+\alpha^2}{\gamma^2}\left[\ddot{\vec{s}}-\left(\vec{s}\cdot\ddot{\vec{s}}\right)\vec{s}\right] +\frac{1}{\mu\gamma}\vec{s}\times\hat{H}\dot{\vec{s}}& =  0
    \end{split}
\end{equation}
supplemented by the boundary conditions defined by the initial and the final orientation of the magnetic moment. Here, $\hat{I}$ is a $3\times 3$ identity matrix and $\hat{H}$ is the matrix of second derivatives of the energy $E$ with respect to components of the magnetic moment $s_x$, $s_y$, $s_z$. Note that Eq.~(\ref{eq:EL}) is derived under the constraint $|\vec{s}|=1$. The optimal switching pulse is found upon substituting the OCP into Eq.~(\ref{eq:three}).

It is not possible to find a general analytical solution to the Euler-Lagrange equation except for special cases where the symmetries of the system make it possible to simplify the problem. For example, for a free magnetic moment ($E=0$) Eq.~(\ref{eq:EL}) simplifies to:
\begin{equation}
    \ddot{\vec{s}}-\left(\vec{s}\cdot\ddot{\vec{s}}\right)\vec{s} = 0,
\end{equation}
and the solution is a constant-speed rotation over the shortest distance between the initial and final states. The corresponding energy cost $\Phi_f$ for reversing of a free macrospin reads
\begin{equation}
    \label{eq:cost_free}
    \Phi_f=\pi^2(1+\alpha^2)/(\gamma^2T).
\end{equation}

Another case with a fully analytical solution is the reversal of a macrospin with uniaxial anisotropy~\cite{kwiatkowski2021optimal}. Because of the rotational symmetry of the problem, the separation of variables in the spherical coordinate system is possible if the $z$-direction is chosen to be along the anisotropy axis. This leads to a well-known Sine-Gordon equation for the polar angle $\theta$ of the magnetic moment and makes the azimuthal angle $\varphi$ completely defined by $\theta$ (see Fig.~\ref{system_picture} for the definition of $\theta$ and $\varphi$):  
\begin{equation}
   \tau_0^2\ddot{\theta} =  \frac{\alpha^2}{4(1+\alpha^2)^2}\sin{4\theta}, \quad \tau_0\dot{\varphi} =  \frac{\cos\theta}{1+\alpha^2},
\end{equation}
where $\tau_0 = \mu/(2\gamma K)$ defines the timescale, and $K$ is the anisotropy constant. Solution of Eq. (10) is explicitly expressed in terms of the Jacobi amplitude~\cite{kwiatkowski2021optimal}; it describes a superposition of the steady rotation of the moment between the energy minima and its precession around the anisotropy axis, where the precession direction reverses when the system reaches the top of the energy barrier. The corresponding optimal switching field rotates synchronously with the magnetic moment in such a way that it generates the torque only in the direction of increasing $\theta$~\cite{kwiatkowski2021optimal}. The amplitude of the optimal switching field remains constant over time when $\alpha=0$, but it exhibits a maximum (minimum) before (after) crossing the energy barrier for $\alpha>0$~\cite{kwiatkowski2021optimal}. The optimal switching field is always perpendicular to the magnetic moment, see Eq.~(4).

Nevertheless, most cases are impossible to solve analytically, and numerical methods for finding OCPs are required. One such method is presented in the following.

\begin{figure}[!t]
\centering
\includegraphics[width=\columnwidth]{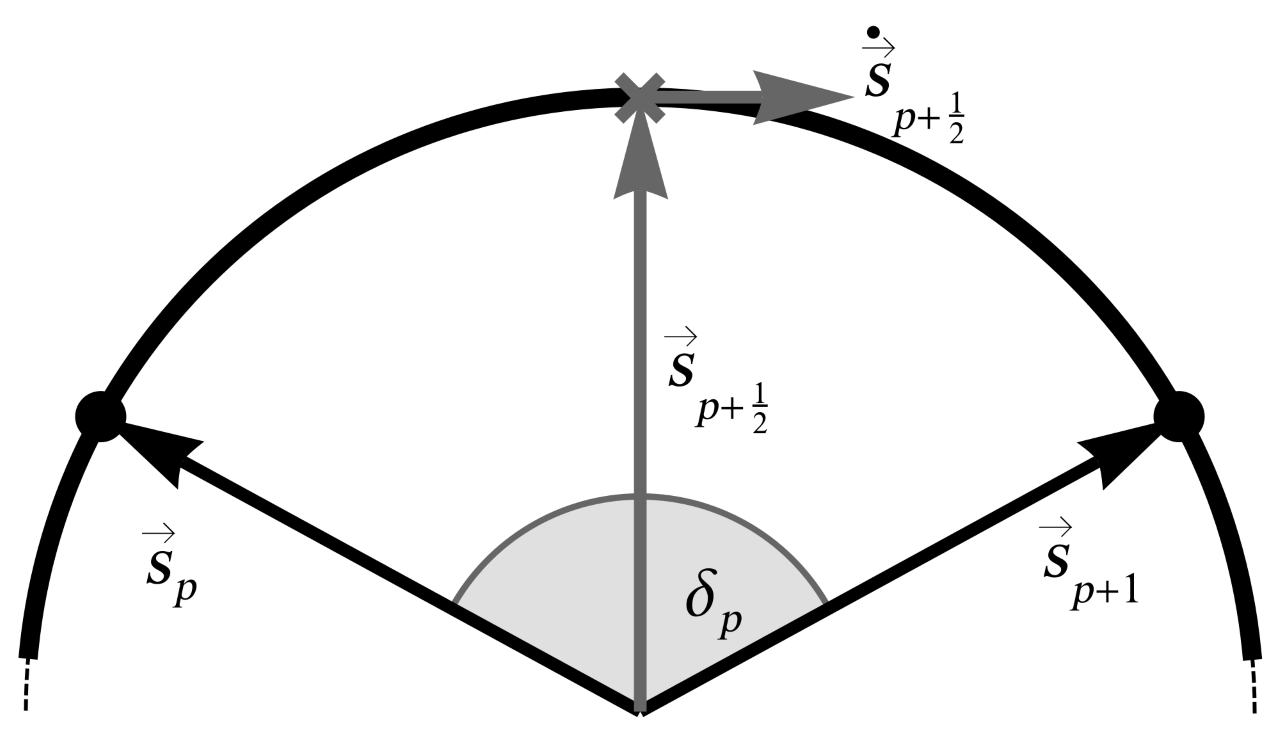}
\caption{\label{numerical}Illustration of the midpoint scheme used in the numerical method for finding OCPs. Two images $\vec{s}_p$ and $\vec{s}_{p+1}$ are connected by a geodesic path in the configuration space. The position $\vec{s}_{p+\frac{1}{2}}$ and the velocity $\dot{\vec{s}}_{p+\frac{1}{2}}$ at the midpoint of the path are defined by $\vec{s}_p$ and $\vec{s}_{p+1}$, and the angle $\delta_p$ between them.} 
\end{figure}

\subsection{Numerical calculation of optimal control paths}
\label{methods:Numerical method for finding OCP}

We find OCPs numerically via the direct minimization of the cost functional. For this, we discretize $\Phi$ using the midpoint rule~\cite{badarneh2020mechanisms}:
\begin{equation} \label{eq:midpoint}
 \Phi[\vec{s}(t)] \approx \Phi[\mathbf{s}]  =  \sum_{p=0}^{Q}  |\vec{B}_{p+\frac{1}{2}}|^2 \left(t_{p+1} - t_p\right),
\end{equation} 
where $\{t_p\}$ is a partition of the time interval $[0,T]$ such that $0=t_0<t_1<\ldots<t_{Q+1}=T$. Here, the partition has a regular spacing, i.e. $t_{p+1} - t_p = \Delta t = T/(Q+1)$, $p=0,\ldots,Q$. A switching trajectory $\vec{s}(t)$ is represented by a polygeodesic line connecting $Q+2$ points, referred to as `images':  $\vec{s}(t)\rightarrow\{\vec{s}_0, \vec{s}_1,...,\vec{s}_{Q+1}\}$, with $\vec{s}_p=\vec{s}(t_p)$. The first image $\vec{s}_0$ and the last image $\vec{s}_{Q+1}$ correspond to the initial and the final orientation of the magnetic moment, respectively; They are fixed, but $Q$ intermediate images can be moved. The external field $\vec{B}_{p+\frac{1}{2}}\equiv \vec{B}(\vec{s}_{p+\frac{1}{2}},\dot{\vec{s}}_{p+\frac{1}{2}})$ is defined by the position and the velocity of the magnetic moment at the midpoint of discretization intervals, see Fig.~\ref{numerical}, via Eq.~(\ref{eq:three}). On the other hand, both $\vec{s}_{p+\frac{1}{2}}$ and $\dot{\vec{s}}_{p+\frac{1}{2}}$ can be expressed in terms of the position of the images:
\begin{align}
    \vec{s}_{p+\frac{1}{2}} &= \frac{ \vec{s}_{p+1}+ \vec{s}_{p}}{|\vec{s}_{p+1}+ \vec{s}_{p}|},\label{eq:position}\\
    \dot{\vec{s}}_{p+\frac{1}{2}} &=\frac{\delta_p}{\Delta t} \frac{ \vec{s}_{p+1}- \vec{s}_{p}}{|\vec{s}_{p+1}- \vec{s}_{p}|},\label{eq:velocity}
\end{align}
where $\delta_p$ is the angle between $\vec{s}_p$ and $\vec{s}_{p+1}$ (see Fig.~\ref{numerical}). Note that the magnitude of $\dot{\vec{s}}_{p+\frac{1}{2}}$ is defined by the finite-difference approximation for the angular velocity, and its direction is along the unit vector $(\vec{s}_{p+1}- \vec{s}_{p})/|\vec{s}_{p+1}- \vec{s}_{p}|$ ensuring orthogonality to $\vec{s}_{p+\frac{1}{2}}$. Upon substituting Eqs.~(\ref{eq:position}), (\ref{eq:velocity}), and (\ref{eq:three}) into Eq.~(\ref{eq:midpoint}), the  functional $\Phi$ becomes a function of a $3Q-$dimensional vector $\mathbf{s}$ defining the position of the movable images, $\mathbf{s}=\left(\vec{s}_1,\ldots,\vec{s}_Q\right)$. 

Possible OCPs of the magnetization switching can be identified by locating minima of $\Phi(\mathbf{s})$. This is done using the velocity projection optimization (VPO) method~\cite{bessarab2015} and/or the limited-memory Broyden-Fletcher-Goldfarb-Shanno (LBFGS) algorithm \cite{nocedal2006numerical, ivanov2021fast} equipped with the force acting on the movable images: 
\begin{equation}
    \mathbf{F}=-\mathbf{\nabla}^\perp\Phi,
    \label{eq:force}
\end{equation}
where $\mathbf{\nabla}^\perp$ denotes the gradient projected on the tangent space of the configuration space, which is a curved manifold due to the constraint $|\vec{s}_p|=1$, $p=1,\ldots,Q$. Explicitly:
\begin{equation}
    \mathbf{\nabla}^\perp=\left(\vec{\nabla}_1-\vec{s}_1(\vec{s}_1\cdot\vec{\nabla}_1),\ldots,\vec{\nabla}_Q-\vec{s}_Q(\vec{s}_Q\cdot\vec{\nabla}_Q)\right),
    \label{eq:proj_grad}
\end{equation}
where $\vec{\nabla}_p\equiv\partial/\partial\vec{s}_p$. For a given number of images involved in the local minimization of $\Phi(\mathbf{s})$, the calculation is considered converged when the magnitude of the force, $|\mathbf{F}|$, has dropped below the set tolerance. 
However, even convergence with a tight force tolerance may be insufficient for a satisfactory resolution of the OCP if $Q$
is not large enough. On the other hand, including too many images in the calculation would result in an unnecessarily high computational effort. Therefore, the following strategy is applied: The OCP search is started with a moderate number of images and the switching path is first converged only to a rather high tolerance so as to bring the images relatively close to the OCP with a reduced computational effort; after that, images are progressively added to the path and minimization of the energy cost is reiterated with a low force tolerance until $\Phi(\mathbf{s})$ stops changing. In this work, up to $Q=1500$ movable images was used depending on the parameters of the system and the switching time, with the lowest force tolerance corresponding to the drop of the force by ten orders of magnitude. 

Some initial arrangement of the images is needed to start an OCP calculation. This can be generated, for example, by placing the images uniformly along the shortest-distance path between the initial and the final state of the transition, or by using some previously found OCP. It is also recommended to add small random noise to the initial path so as to avoid convergence on maxima or saddle points of $\Phi(\mathbf{s})$ due to possible symmetries in the system. A local minimization of $\Phi(\mathbf{s})$ will most likely converge to the OCP closest to the initially generated path. If multiple OCPs are present between the initial and the final state, several initial estimates need to be produced so as to enable convergence on the various solutions. 

\begin{figure*}[!t]
\centering
\includegraphics[width=\textwidth]{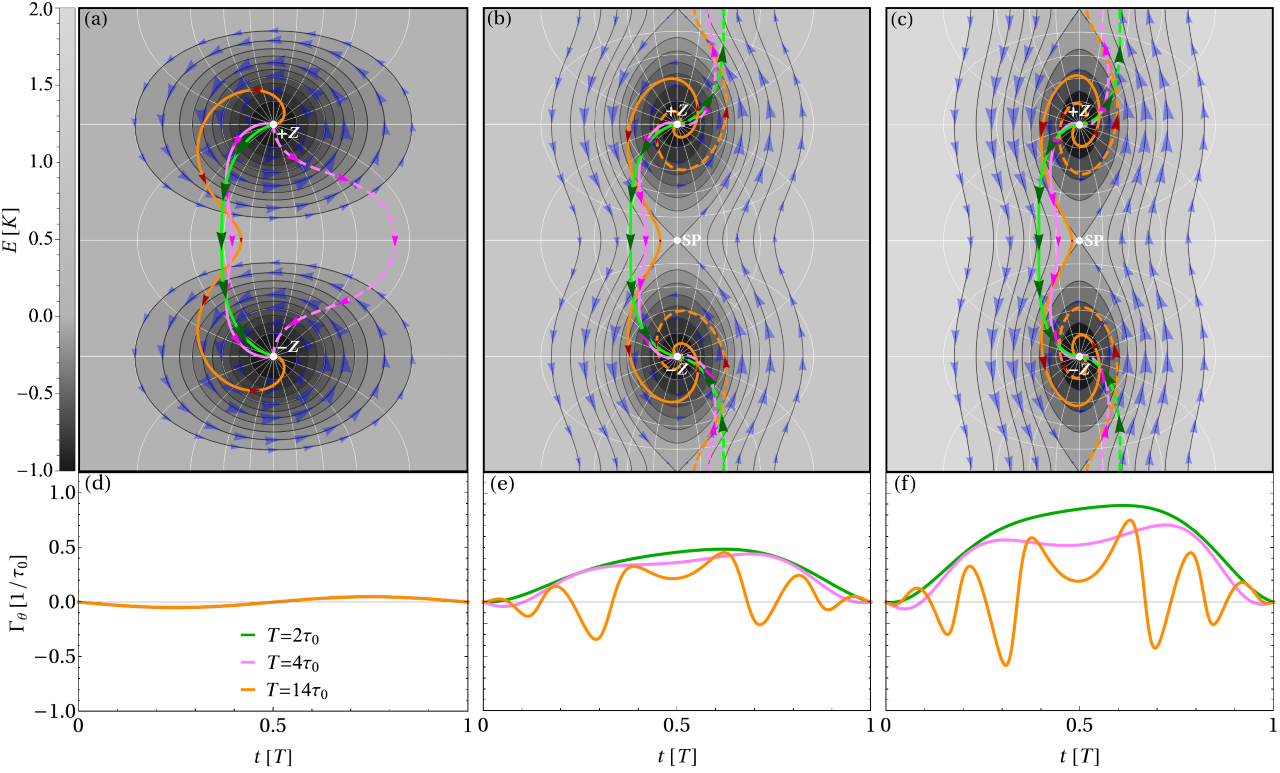}
\caption{\label{contour}Transverse Mercator projection~\cite{deetz1945elements} of the energy surface of a macrospin with (a) uniaxial anisotropy and biaxial anisotropy with (b) $\xi=1$  and (c) $\xi=2$. The meridians and the parallels (see Fig.~\ref{system_picture}) are shown with thin white lines. The blue arrows show the distribution of the internal torque, with the size of the arrows being proportional to the magnitude of the torque. The calculated OCPs between the energy minima at $+Z$ and $-Z$ are shown with the green, pink, and orange lines for $T = 2\tau_0$, $T=4\tau_0$, and $T=14 \tau_0$, respectively. The arrows along each OCP show the velocity at $t=T/6$, $t=T/3$, $t=T/2$, $t=2T/3$, and $t=5T/6$, where the arrow size codes the magnitude of the velocity. The damping factor $\alpha$ is $0.1$. The solid and the dashed lines of the same color show equivalent OCPs. They differ by an arbitrary rotation around the easy axis for the uniaxial case; For finite $\xi$, the degeneracy is lifted and there are two OCPs, symmetrical with respect to a $\pi$-angle rotation around the easy axis, for a given $T$. Note that the OCPs do not pass through saddle points (SP) on the energy surface. The $\theta$-projection of the internal torque along the OCPs from (a)-(c) are shown in (d)-(f), respectively.
}
\end{figure*}

\subsection{Perturbation theory}
\label{methods:Perturbation theory}

Although it is not possible to obtain a general analytical solution to the Euler-Lagrange equation [see Eq.~(\ref{eq:EL})], some analytical estimates for the energy-efficient switching can still be derived using perturbation theory. For this, we expand Eq.~(\ref{eq:EL}) around the free-macrospin solution and obtain the OCP in terms of perturbation series with respect to the parameters defined by the Hamiltonian of the system (see Appendix~\ref{appendix} for details). The minimum energy cost of switching can be estimated based on the approximate solution for the OCP. The second-order expansion for $\Phi_m$ is used in particular:
\begin{equation} \label{eq:cost functional approximation}
  \Phi_m \approx \Phi_f + \sum_{i=1}^N\epsilon_i \Phi_i + \sum_{i,j=1}^N\epsilon_i \Phi_{ij} \epsilon_j,
\end{equation}
where $N$ is the number of independent perturbations, $\epsilon_i$ is the $i$th dimensionless perturbation parameter, and $\Phi_i$, $\Phi_{ij}$ are the expansion coefficients describing the first- and the second-order corrections, respectively. The explicit expressions for $\epsilon_i$, $\Phi_i$, and $\Phi_{ij}$ for the biaxial system are presented in  the following.

\section{Results}
\label{results}
Here, we apply the methodology presented earlier to the magnetization reversal in a biaxial anisotropy system, e.g. to a flat elongated nanomagnet shown in Fig.~\ref{system_picture}. The internal energy of the system is given by the following equation 
\begin{equation} \label{energy_biaxial}
    E = \xi K s_x^2 -K s_z^2,
\end{equation}
where the easy axis and the hard axis are along the $z$ and $x$ directions, respectively, $K>0$ is the anisotropy constant, and $\xi$ is a dimensionless parameter defining the relative strength of the hard-axis anisotropy. The energy surface of the system has two minima at $\vec{s}=(0,0,1)$ and $\vec{s}=(0,0,-1)$, and two saddle points at $\vec{s}=(0,1,0)$ and $\vec{s}=(0,-1,0)$ (see Fig.~\ref{system_picture}). This model is commonly used to describe in-plane memory bits~\cite{chun2012scaling} and elements of artificial spin ice systems~\cite{nisoli2013colloquium,wysin2012magnetic}. Energy-efficient switching between the energy minima in time $T$ is analysed in the following. 

\subsection{Optimal protocols for magnetization reversal}
Figures \ref{contour}(a)-(c) show the calculated OCPs of the magnetization reversal for $\alpha=0.1$ and various switching times and strengths of the hard-axis anisotropy, superimposed on the energy surface of the system. For short switching time, i.e., when $T\sim \tau_0$, the OCPs deviate weakly from geodesic paths between the energy minima. With increasing $T$, the OCPs acquire precessional motion around the easy axis, where the sense of precession changes upon reaching the top of the energy barrier at $s_z=0$.

\begin{figure}[!t]
\centering
\includegraphics[width=\columnwidth]{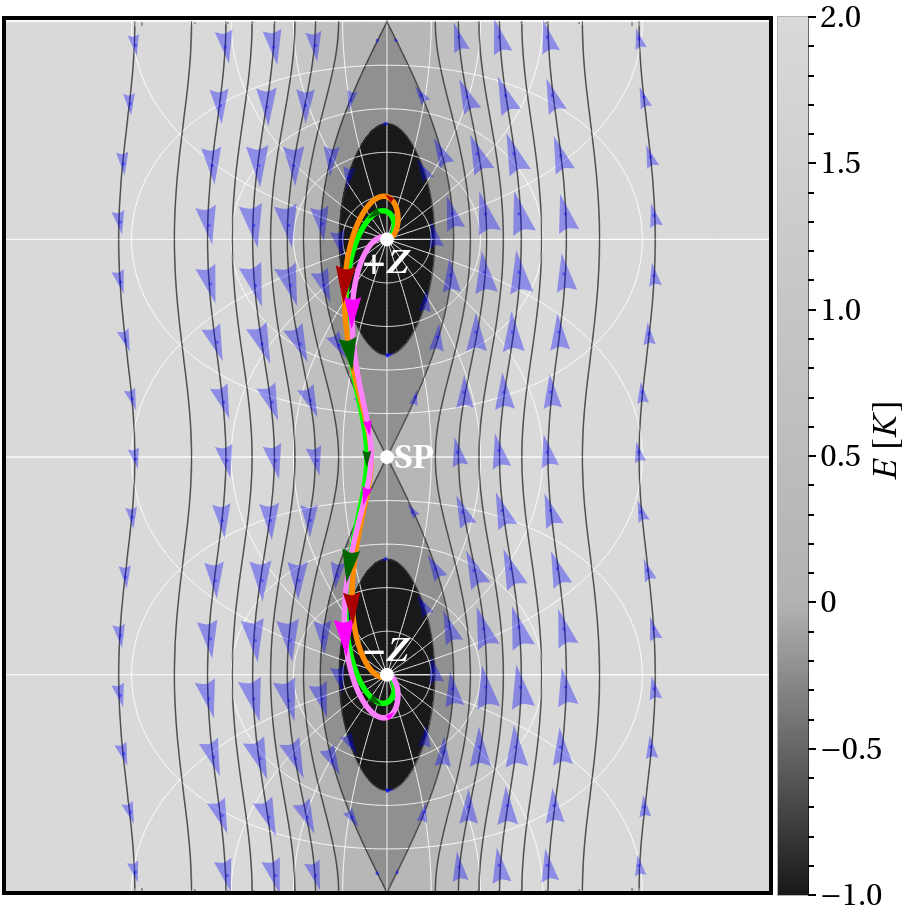}
\caption{\label{asymm_OCPs}Calculated OCPs for $\xi=4$, $\alpha=0.2$, and $T=5.314\tau_0$. The notations are the same as in Fig.~\ref{contour}(a)-(c). The OCP shown with the green line exhibits a mirror symmetry with respect to the $XY$ plane, but the symmetry is broken for the OCPs shown with pink and orange lines. The asymmetric OCPs correspond to the global minimum of $\Phi$ for the given parameter values. Note that the asymmetric OCPs can be
obtained from one another via reflection in the $XY$ plane. More OCPs can be obtained by a $\pi$-angle rotation of the shown OCPs around the easy axis. There are in total six OCPs in the present case.}
\end{figure}

The $\xi=0$ case describes a uniaxial-anisotropy system, for which OCPs can be found analytically~\cite{kwiatkowski2021optimal}. Due to the axial symmetry, the OCPs for a fixed switching time are degenerate with respect to overall rotation around the easy axis. For example, the two OCPs shown by the solid and dashed pink lines in Fig.~\ref{contour}(a) are equivalent. In contrast, the axial symmetry is broken when $\xi\neq 0$, which results in the well-separated OCPs between the energy minima. In most cases, there are two equivalent, mirror-symmetric (with respect to the $XY$ plane) OCPs in the biaxial system~\cite{barros2013microwave} for a given switching time, and the OCPs differ by a $\pi$-angle rotation around the easy axis, see Fig.~\ref{contour}(b)-(c) and also Fig.~\ref{system_picture}. However, more co-existing OCPs can be present for $\xi\gtrsim4$, where the paths are different by the amount of precession around the initial and the final states (see Fig. \ref{asymm_OCPs}). Note that the OCPs can break the $XY$-plane mirror symmetry. For certain parameter values, such asymmetric OCPs deliver the global minimum to the functional $\Phi$, which is the case shown in Fig. \ref{asymm_OCPs}, or even represent the only type of solution. Nevertheless, the OCPs never pass through saddle points (SPs) on the energy surface, therefore the system does not cross the lowest possible energy barrier within the energy efficient switching protocol. 

\begin{figure}[!t]
\centering
\includegraphics[width=\columnwidth]{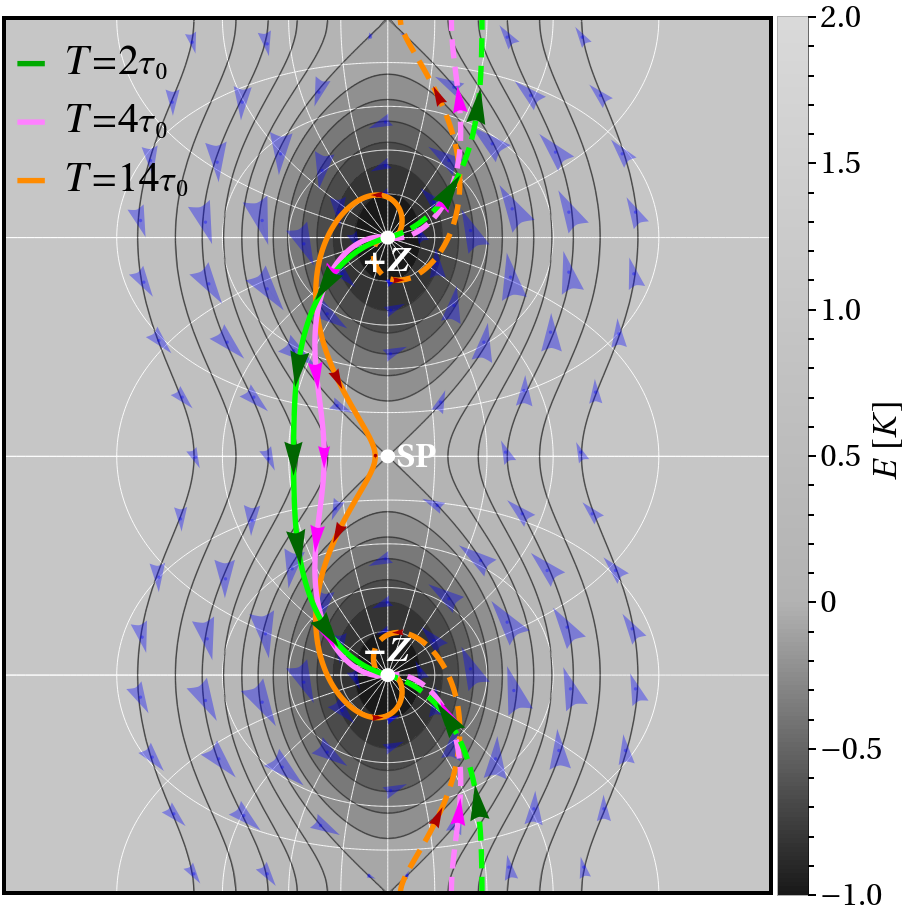}
\caption{\label{contour_xi1}Calculated OCPs for $\xi=1$, $\alpha=0.4$, and several values of the switching time, as indicated in the legend. The notations are the same as in Fig.~\ref{contour}(a)-(c).}
\end{figure}

\begin{figure*}[!t]
\centering
\includegraphics[width=\textwidth]{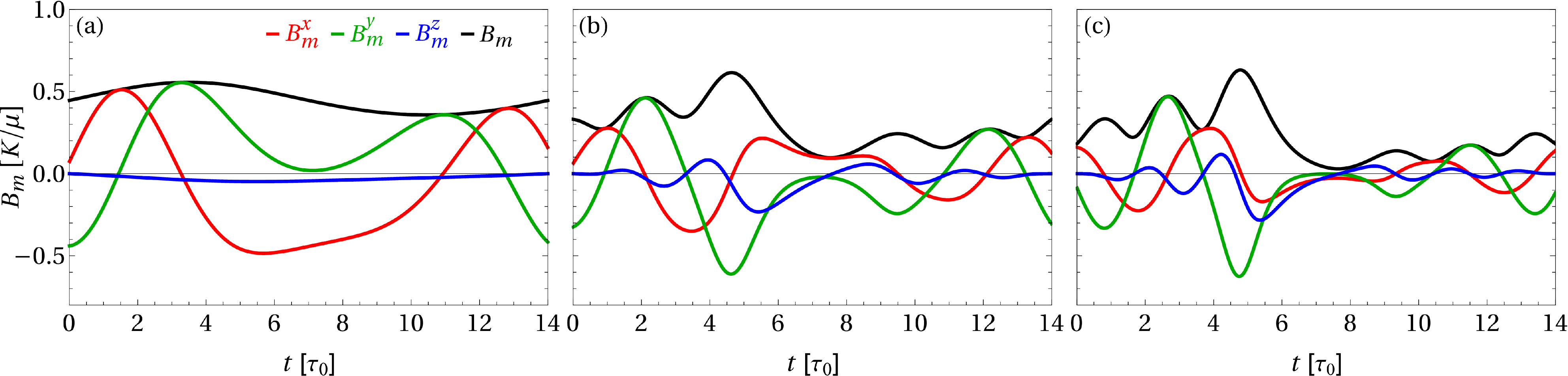}
\caption{\label{optimal_field}Calculated optimal switching pulse of external magnetic field for a macrospin with (a) uniaxial anisotropy and biaxial anisotropy with (b) $\xi=1$ and (c) $\xi=2$. The switching time $T$ is $14\tau_0$ and the damping parameter $\alpha$ is 0.1. The pulses are derived from the OCPs shown in Fig.~\ref{contour}(a)-(c).
}
\end{figure*}

The distribution of the internal torque $\vec{\Gamma}$ [see the blue arrows in Fig.~\ref{contour}(a)-(c)] provides an insight into the mechanism of energy-efficient magnetization switching in biaxial systems and explains the position and shape of calculated OCPs. When $\xi=0$, the torque only generates precession around the easy axis and, in case of nonzero damping, relaxation to the energy minima. In this case, the internal torque does not assist switching since it does not point in the direction of the final state anywhere in the region of the initial state ($s_z>0$). This behavior is described quantitatively by the component of $\vec{\Gamma}$ in the direction of increasing $\theta$, relevant for the reversal process:
\begin{equation} \label{internal_torque}
 \Gamma_{\theta} = \Gamma_0 \sin \theta \left[\xi \sin(2\varphi)-2\alpha \cos \theta (1+\xi \cos^2 \varphi)\right],  
\end{equation}
where $\Gamma_0 \equiv \left[2\tau_0 (1+\alpha^2)\right]^{-1}$. $\Gamma_\theta$ along the calculated OCPs is shown in Fig.~\ref{contour}(d)-(f). Positive (negative) $\Gamma_\theta$ signifies positive (negative) contribution of the internal torque to the reversal. For $\xi=0$, Eq.~(\ref{internal_torque}) reduces to $\left.\Gamma_{\theta}\right|_{\xi=0} = -\alpha\Gamma_0\sin 2\theta$. Clearly, $\left.\Gamma_{\theta}\right|_{\xi=0}<0$ for $\theta<\pi/2$ and nonzero $\alpha$ [see Fig.~\ref{contour}(d)]. 

Adding a hard-axis anisotropy to the system for ($\xi>0$) gives the contribution to the internal torque in the switching direction in a certain sector of the configuration space, see Fig.~\ref{contour}(b)-(c). The location of the calculated OCPs in this sector demonstrates the principle of energy-efficient control which lies in the effective use of the system's internal dynamics. It is now clear why the OCPs do not pass through the SP, where the internal torque vanishes: It is beneficial to climb up the energy surface where the internal torque picks up and assists the switching process. In particular, $\Gamma_\theta$ is maximized at the equator ($\theta=\pi/2$) when $\varphi=\pi/4$ and $\varphi=5\pi/4$  [see Eq.~(\ref{internal_torque})]. In an optimal protocol, a balance is reached between the effort in climbing up the energy surface and the strength of the internal torque. As a result, the OCPs cross the equator at an optimal point $\pi/4<\varphi_m<\pi/2$ or $5\pi/4<\varphi_m<3\pi/2$, see Fig.~\ref{contour} (b)-(c). 

The favorable effect of the torque produced by the hard axis is also evident from the $\Gamma_\theta(t)$ dependencies calculated along the OCPs, see Fig.~\ref{contour}(e)-(f). Although there are regions where $\Gamma_\theta<0$, the $\theta$-component of the torque is mostly positive, especially for shorter switching times, and the magnitude of the torque increases with $\xi$. It is noteworthy that the asymmetric shape of $\Gamma_{\theta}$ about $T/2$ -- the result of the damping contribution to the torque -- does not contradict to the mirror-symmetry of the OCPs. For symmetric OCPs, the total torque stays symmetric.

Figure \ref{contour_xi1} shows the calculated OCPs for $\xi=1$ and $\alpha=0.4$. The OCPs look similar to those calculated for weaker damping, but demonstrate less precession, which is particularly seen for longer switching time. Overall, increased damping makes the internal torque deviate stronger from the energy contours toward the energy minima, cf. Fig.~\ref{contour}(b), leading to an increase in the energy cost of switching. This effect is analysed quantitatively in the following section.

The optimal switching pulses of external magnetic field for $T=14\tau_0$, $\alpha=0.1$, and $\xi=0,1,2$ are presented in Fig. \ref{optimal_field}. Note that the pulses are derived from the OCPs presented in Fig. \ref{contour} using Eq.~(\ref{eq:three}).  As expected, increasing strength of the hard-axis anisotropy leads to overall decrease in the field amplitude, although its peak values can exceed the maximum field value in the $\xi=0$ case. 

The experimental realization of optimal control pulses is a challenging task but still feasible within current pulse shaping technology~\cite{gerrits2002ultrafast,mayer2007microstrip,curcic2008polarization,gao2013simple,bisig2015dynamic,rius2016incoherent}. It is worth noting that the optimal switching protocols remain quite stable with respect to thermal fluctuations and material parameter perturbations, as confirmed by our spin dynamics simulations (see Appendix B).

Similarly to the uniaxial case, the pulse is stronger in the first half of the reversal where relaxation works against the switching process, and weaker in the second half where relaxation pushes the system to the desired energy minimum. However, there is a distinct oscillation in the field amplitude associated with the broken axial symmetry of the system. This amplitude oscillation is present even at zero damping, which is in contrast to the uniaxial case where the field amplitude is time-independent for $\alpha=0$~\cite{kwiatkowski2021optimal}. The amplitude peaks when the magnetic moment deviates most from the easy plane where the energy gradient and, thereby, the internal torque are the largest. Furthermore, the external pulse amplitude is the lowest close to $t=T/2$, where the internal torque brings the system over the barrier. Irrespective of the $\xi$ value, the switching field is always perpendicular to the magnetic moment [see Eq.~(4)] and its amplitude $B_m$ demonstrates the symmetry: $B_m(0) = B_m(T)$. Note that $B_m(0) = B_m(T/2) = B_m(T)$ in the $\xi=0$ case~\cite{kwiatkowski2021optimal}.

\subsection{Minimum energy cost of switching}

\begin{figure*}[!t]
\centering
\includegraphics[width=\textwidth]{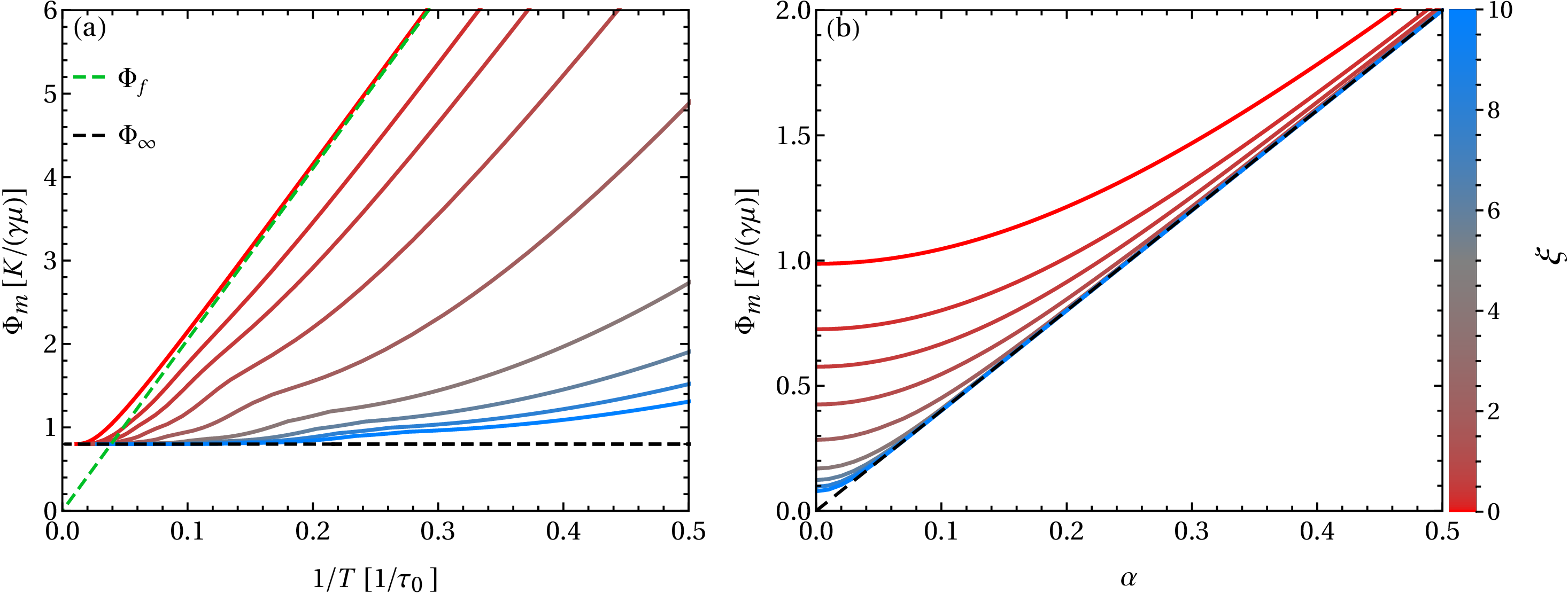} 
\caption{Minimum energy cost of magnetization reversal as a function of (a) inverse of the switching time for $\alpha=0.2$, (b) damping parameter for $T = 20 \tau_0$, for various $\xi$ values. Green dashed line corresponds to the solution of the reversal of a free macrospin, while the black dashhed line shows the infinite switching time asymptotic.
} 
\label{fun_vs_invrT_and_al}
\end{figure*}

The revealed optimal reversal protocols can now be used to calculate the minimum energy cost of switching $\Phi_m$ using Eq.~(\ref{eq:midpoint}) (see also Ref.~\cite{kwiatkowski2021optimal} for the analytical solution for the $\xi=0$ case). In the following, we always pick the lowest value of the energy cost whenever multiple OCPs are present for a given set of parameters. Figure \ref{fun_vs_invrT_and_al}(a) shows $\Phi_m$ as a function of the inverse of the switching time for $\alpha=0.2$ and various strengths of the hard-axis anisotropy. For any $\xi$ value, 
$\Phi_m$ decreases monotonically with $T$ and approaches the universal lower limit 
\begin{equation}
\label{eq:lower_limit}
\Phi_{\infty} \equiv 4 \alpha K/ \left(\gamma \mu\right)
\end{equation}
at infinitely long switching time~\cite{barros2013microwave}. Note that $\Phi_m$ reaches the limit faster for larger values of $\xi$. Overall, there is a decrease in $\Phi_m$ with $\xi$, as expected from the distribution of the torque in biaxial systems. 

The switching cost for a free macrospin $\Phi_f(T)$ (see the green dashed line in Fig.~\ref{fun_vs_invrT_and_al}(a)) provides a useful benchmark for evaluating the favorable effect of the torque produced by the hard axis. Notably, the switching cost can be significantly lower than $\Phi_f(T)$ in a certain range of $T$ for finite strengths of the hard-axis anisotropy. For example, $\Phi_m(T)$ becomes almost an order of magnitude smaller than $\Phi_f(T)$ for $\xi=10$ and $T\approx 2\tau_0$. This is in contrast to the uniaxial-anisotropy case ($\xi=0$), where $\Phi_m(T)\ge\Phi_f(T)$ (the equality is reached for $\alpha=0$) for any given $T$.

The $\alpha$-dependencies of the minimum switching cost for $T=20\tau_0$ and several values of $\xi$ are shown in Fig.~\ref{fun_vs_invrT_and_al}(b). Irrespective of the strength of the hard-axis anisotropy, $\Phi_m$ is a monotonically increasing function of the damping parameter, approaching the $\Phi_\infty$ asymptote when $\alpha\rightarrow\infty$. It is noteworthy that the reduction in the switching cost with $\xi$ becomes more pronounced as $\alpha$ decreases. 

\begin{figure*}[!t]
\centering
\includegraphics[width=\textwidth]{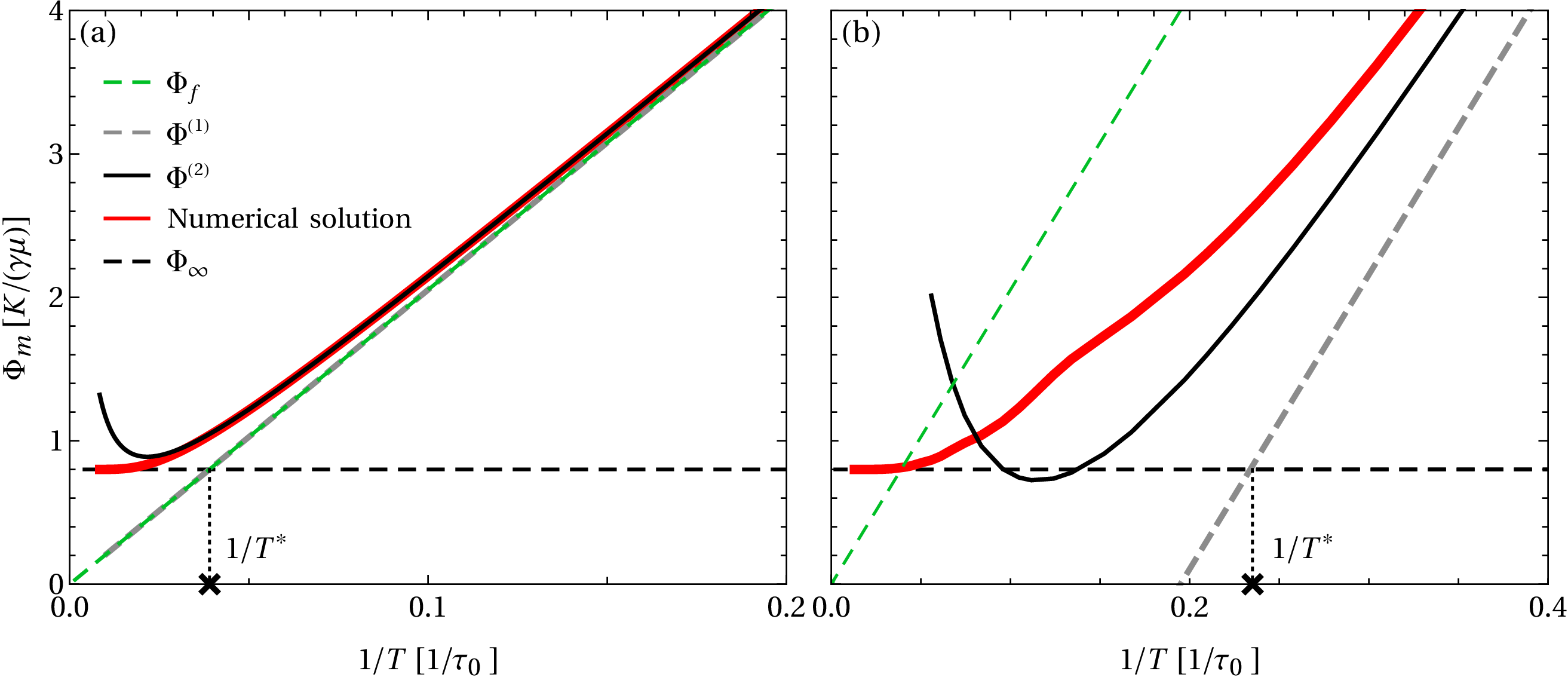} 
\caption{\label{phi_vs_InvrsT_all}Approximation for the minimum energy cost of magnetization reversal for a macrospin within the zero- ($\Phi_f$), first- ($\Phi^{(1)}$), and second-order ($\Phi^{(2)}$) perturbation theory [see Eq. (\ref{min_func_biaxial})], as indicated in the legend, vs the inverse of the switching time. The strength of the hard-axis anisotropy $\xi$ is (a) 0 and (b) 1. Red solid line shows the numerically exact solution. Black dashed line shows the infinite switching time asymptotic. The intersection of the short and the long switching time asymptotes provides the optimal switching time $T^\ast$ [see Eq. (\ref{optimal_time})]. The magnitude of $\alpha$ is $0.2$.} 
\end{figure*}

\subsection{Perturbation theory analysis} \label{sec:perturbation}

Both anisotropies in the biaxial system can be treated as independent perturbations to the free macrospin. This results in two dimensionless perturbation parameters $\epsilon_1 \equiv \xi T/\tau_0$ and $\epsilon_2 \equiv T/\tau_0$ defined by the hard- and the easy-axis anisotropy, respectively. The approximation to the minimum energy cost is obtained by substituting the perturbation series for the OCP (see Appendix~\ref{appendix}) into Eq.~(\ref{eq:cost_funct}). The result, up to the second-order terms, reads 
\begin{equation} \label{min_func_biaxial}
  \Phi_m \approx \Phi_f-\frac{4K}{\gamma\mu}\xi +\frac{K^2T}{2(1+\alpha^2)\mu^2}\left[\alpha^2+\alpha^2\xi+\frac{1}{4}(4+5\alpha^2)\xi^2\right],  
 \end{equation}
where the contributions from the hard-axis anisotropy are recognized by the $\xi$-factor. The smallness of the perturbation parameters, $\epsilon_1,\epsilon_2\ll 1$, can be translated into the condition on $T$: $T\ll \tau_0$. Therefore, Eq.~(\ref{min_func_biaxial}) can be interpreted as a short switching time approximation for $\Phi_m$. 

In Eq. (\ref{min_func_biaxial}), the first and the second terms represent the free macrospin solution and the first-order correction, respectively; the rest of the equation describes the second-order correction that includes the terms due to the easy- and the hard-axis anisotropies, as well as the cross term.

Equation~(\ref{min_func_biaxial}) clearly shows that the switching cost reduction in biaxial magnets is captured within the linear response to the hard-axis anisotropy. Note that the easy axis does not contribute to the first-order correction; It can be shown in fact that all odd-order corrections vanish in the uniaxial case. 
The approximation for $\Phi_m$ within zero-, first-, and second-order perturbation theory is shown in Fig.~\ref{phi_vs_InvrsT_all} for $\alpha=0.2$. The numerically exact solution for $\Phi_m$ is also shown for comparison. The short switching time approximation eventually breaks down as $T$ increases, and $\Phi_m$ converges on $\Phi_\infty$.

The minimum energy cost for switching has two clear asymptotics: $\Phi_m=\Phi_f - 4K\xi/(\gamma \mu)$ when $T\rightarrow 0$, and $\Phi_m=\Phi_\infty$ when $T\rightarrow \infty$. Their intersection point
\begin{equation}\label{optimal_time}
    T^\ast= \frac{\left(1+\alpha^2\right) \pi^2}{2 \left(\alpha+\xi\right)}\tau_0,
\end{equation}
can be interpreted as an optimal switching time in a sense that increase in $T$ beyond $T^\ast$ does not lead to a significant reduction in the energy cost. Therefore, $T^\ast$ provides a tradeoff between the switching speed and energy efficiency~\cite{kwiatkowski2021optimal,Vlasov2022}. Note that $T^\ast$ decreases with increasing strength of the hard-axis anisotropy.

\section{Conclusions and discussion}
\label{conclusion}

In conclusion, we explored by means of the optimal control theory energy-efficient protocols for magnetization reversal in biaxial nanomagnets. We calculated OCPs of the reversal and used them to derive optimal switching pulses of external magnetic field. We studied the energy cost of switching as a function of the system parameters and the switching time. The internal torque produced by the hard-axis anisotropy can significantly reduce the switching cost: For a given switching time, it can drop below what is needed to reverse a free macrospin, which is impossible in uniaxial-anisotropy systems. However, the energy cost can never be smaller than a universal lower limit defined by the energy barrier and damping~\cite{barros2013microwave}.

We obtained some analytical estimates regarding the reduction of the energy cost using perturbation theory. In particular, we identified the optimal switching time providing a tradeoff between the switching speed and energy efficiency. The optimal switching time decreases with the strength of the hard-axis anisotropy.

\begin{figure}[!t]
\centering
\includegraphics[width=\columnwidth]{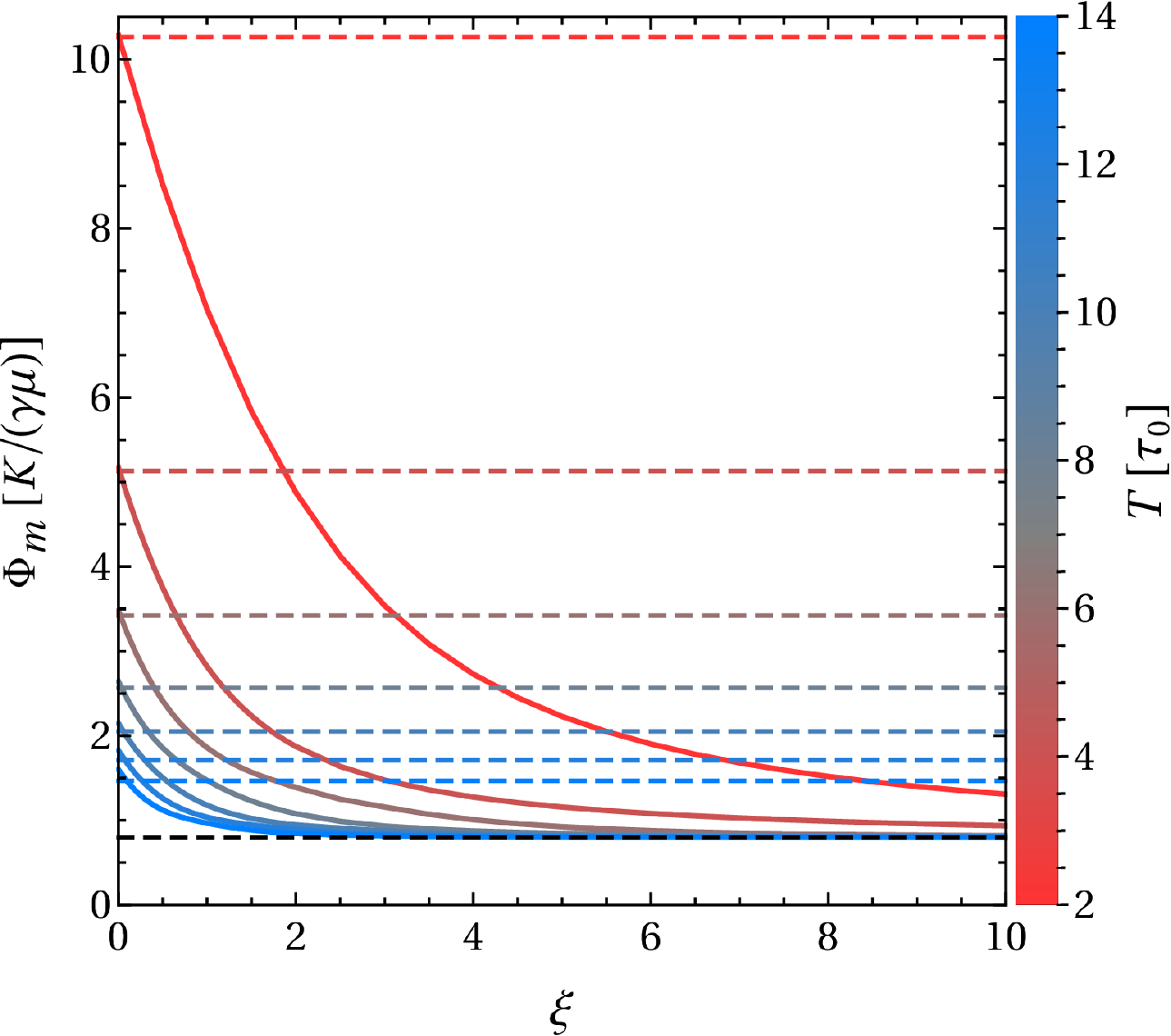}
\caption{\label{func_vs_xi}Minimum energy cost of magnetization reversal as a function of $\xi$ for various $T$ values (solid lines). The dashed color lines show the switching cost for a free macrospin. The magnitude of the damping factor $\alpha$ is $0.2$. Black dashed line shows the infinite switching time asymptotic. 
}
\end{figure}

It is important to realize that the decrease in the switching cost can be achieved in biaxial magnets without sacrificing their thermal stability. Indeed, the thermal stability is characterized by the energy barrier separating the stable states. Within harmonic rate theories, this is defined by the energy difference between the saddle point and the initial state minimum \cite{bessarab2012harmonic,Fiedler2012}. In a biaxial system, the energy barrier amounts to $K$ irrespective of the $\xi$ value, see Eq.~(\ref{energy_biaxial}) and the text around it. In contrast, $\Phi_m$ depends strongly on $\xi$, especially for short switching times, which is particularly clear from Fig.~\ref{func_vs_xi} (note that $\Phi_m$ converges to $\Phi_\infty$ for $\xi\rightarrow\infty$ irrespective of the switching time). The possibility to independently maximize both writability and thermal stability of biaxial magnets makes these systems efficient memory elements that provide a solution to the magnetic recording dilemma.

A macrospin approximation is used in the present study, but this is expected to break down with increasing system size. Even if the initial and the final states are collinear, the transition between them may involve non-uniform rotation of magnetization such as nucleation and propagation of domain walls~\cite{Braun1993,Braun1994,bode2004shape,krause2009magnetization,bessarab2013size} or excitation of spin waves \cite{seki2013spin,Spinelli2014,Yamaji2018, badarneh2020mechanisms}. It remains to be seen under what conditions these and possibly other, yet unknown switching mechanisms become optimal in terms of energy efficiency. This is a subject of future study.

\begin{acknowledgments}
The authors would like to thank B. Hj\"orvarsson, V. Kapaklis, K.A. Th\'orarinsd\'ottir, T. Sigurj\'onsd\'ottir for helpful discussions and valuable comments. This work was supported by the Icelandic Research Fund (Grant Nos. 217813 and 184949), the University of Iceland Research Fund (Grant No. 15673), and the Swedish Research Council (Grant No. 2020-05110).
\end{acknowledgments}
 
\appendix
\section{Approximate solution for optimal control path}\label{appendix}

The Euler-Lagrange equation [see Eq.~(\ref{eq:EL})] in spherical coordinates $\theta$ and $\varphi$ reads

\begin{equation} \label{eq:euler_theta_phi_biaxial}
\begin{split}
 \ddot{\theta} & =  A_0\dot{\varphi}^2
+A_1\dot{\varphi}+A_2, \\
 \ddot{\varphi} & =  C_0\dot{\theta}\dot{\varphi}+C_1\dot{\theta}+C_2.
 \end{split}
\end{equation}
for a biaxial system whose energy is defined by Eq. (\ref{energy_biaxial}), the coefficients become 
\begin{widetext}
\begin{equation}
\label{Coefficient_biaxial}
\begin{split}
A_0 &= \frac{\sin2\theta}{2}, \hspace{0.1cm}
A_1 = \frac{ (2 +\xi) (\sin \theta -3 \sin3\theta)}{8 (1+\alpha^2) \tau_0}  + \frac{3\xi \cos2\varphi \sin^3 \theta}{2 (1+\alpha^2) \tau_0}, \hspace{0.1cm}
A_2 =\frac{\sin 4 \theta (2 +\xi \cos 2 \varphi+\xi )^2}{16 (1+\alpha^2) \tau_0^2} +\frac{ \xi^2 \sin^22 \varphi \sin 2 \theta}{8 (1+\alpha^2) \tau_0^2},\\ 
C_0 &= -2\cot \theta, \hspace{0.1cm}
C_1 = \frac{ (2 +\xi) (3 \cos 2 \theta+1) \csc \theta}{4 (1+\alpha^2) \tau_0}-\frac{3 \xi \cos 2 \varphi \sin \theta}{2 (1+\alpha^2) \tau_0}, \hspace{0.1cm}
C_2 = \frac{ -\xi (2 +\xi) \sin 2 \varphi \cos^2\theta}{2(1+\alpha^2) \tau_0^2}+\frac{\xi^2  \sin4 \varphi \sin^2 \theta}{4(1+\alpha^2) \tau_0^2}.
\end{split}
\end{equation}
\end{widetext}
We seek for $\theta_m(t)$ and $\varphi_m(t)$ -- the 
solution of Eq. (\ref{eq:euler_theta_phi_biaxial}) -- in a form of a series in the two perturbation parameters $\epsilon_1$ and $\epsilon_2$ defined by the biaxial anisotropy (see Section~\ref{sec:perturbation}). In particular, the second-order expansion for $\theta_m(t)$ and $\varphi_m(t)$ reads 
\begin{equation}  \label{eq:theta_phi_series_biaxial}
\begin{split}
      \theta_m(t) &\approx \theta_f(t) + \sum_{i=1}^2\epsilon_i \theta_i(t) +  \sum_{i,j=1}^2\epsilon_i \theta_{ij}(t) \epsilon_j,\\
    \varphi_m(t) &\approx \varphi_f(t) + \sum_{i=1}^2 \epsilon_i \varphi_i(t) +  \sum_{i,j=1}^2\epsilon_i \varphi_{ij}(t) \epsilon_j.
\end{split}
\end{equation}
Here,  
$\theta_f(t)\equiv \pi t/T$ and $\varphi_f(t)\equiv \pi/4,5\pi/4$ describe the reversal of a free macrospin, and the coefficients $\theta_i(t)$, $\varphi_i(t)$, $\theta_{ij}(t)$, $\varphi_{ij}(t)$ are obtained upon substituting Eqs.~(\ref{eq:theta_phi_series_biaxial}) into Eq.~(\ref{eq:euler_theta_phi_biaxial}) and collecting terms with equal powers of $\epsilon_1$ and $\epsilon_2$, which gives the following result 
\begin{widetext}
\begin{equation}
\begin{split}
    \theta_{1} & = 0, \hspace{0.1cm}
    \theta_{2} = 0, \hspace{0.1cm}
    \varphi_{22}  =  0, \hspace{0.1cm}
    \theta_{11} = \frac{\sin\left(2\pi t/T\right)\left[4+4\alpha^2+\alpha^2 \cos\left(2\pi t/T\right)\right]}{128\pi^2 \left(1+\alpha^2\right)^2}, \\
    \theta_{12} &= \theta_{21} = -\frac{\alpha^2 \sin \left(4 \pi  t/T\right)}{128 \pi ^2 \left(1+\alpha^2\right)^2 }, \hspace{0.1cm}
    \theta_{22} = 2\theta_{12}, \hspace{0.1cm}
    \varphi_1 = -\frac{\left(8-\alpha^2\right) }{64 \left(1+\alpha^2\right)} + \frac{  \sin \left(\pi  t/T\right)}{2 \pi \left(1+\alpha^2\right)}, \\
    \varphi_2 &= 2\varphi_1, \hspace{0.1cm}
    \varphi_{11} = \varphi_{12} = \varphi_{21} =  \frac{\left(4+\alpha^2\right)  \cos \left(2 \pi  t/T\right)}{32 \pi ^2 \left(1+\alpha^2\right)^2 } - \frac{3\left(8-\alpha^2\right)}{2048 \left(1+\alpha^2\right)} + \frac{100+73\alpha^2}{480\pi^2 (1+\alpha^2)^2}. 
\end{split}
\end{equation}
\end{widetext}
The approximation for the minimum energy cost of switching, Eq.~(\ref{min_func_biaxial}), is obtained upon substituting Eq.~(\ref{eq:theta_phi_series_biaxial}) into Eq.~(\ref{eq:cost_funct}).

\begin{figure*}[t!]
\centering
\includegraphics[width=\textwidth]{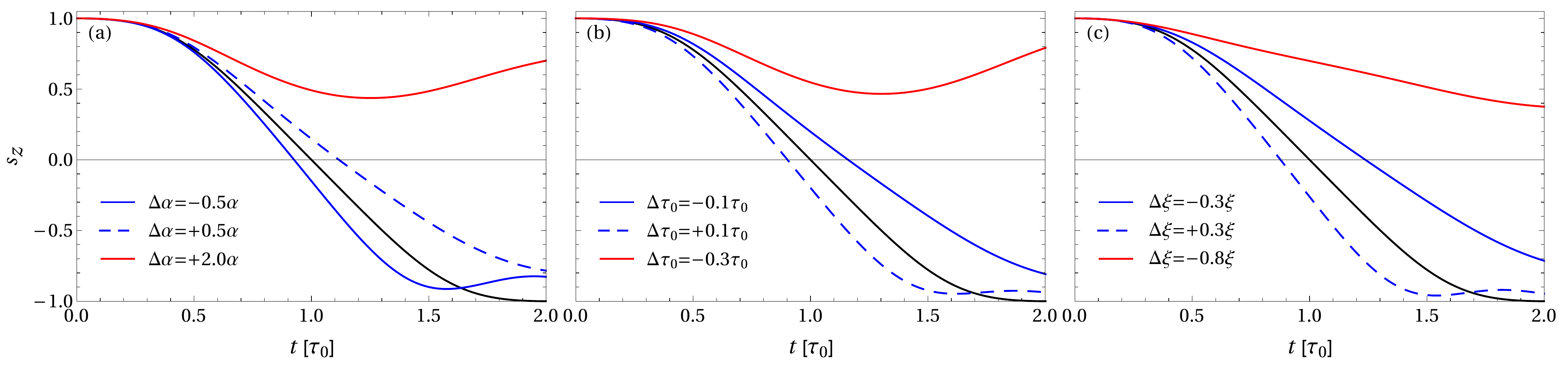}
\caption{Effect of perturbations in the material parameters $\alpha$ (a), $\tau_0$ (b), and $\xi$ (c) on the magnetization reversal induced by the optimal switching pulse. Magnitude of the perturbations $\Delta \tau_0$, $\Delta \alpha$, and $\Delta \xi$ is shown in the legend. Blue (red) lines show evolution of the z-component of the normalized magnetic moment during successful (unsuccessful) reversal. Black line corresponds to the reversal in a particle characterized by unperturbed material parameters: $\alpha=0.1$, $\xi=5$.}
\label{fig10}
\end{figure*}

\section{Spin dynamics simulations}

The robustness of the optimal switching protocol for the biaxial monodomain particle [See Eq.~(\ref{energy_biaxial})] against thermal fluctuations and perturbations in the material parameters was tested by carrying out additional spin dynamics simulations. The simulations involved time integration of the Landau-Lifshitz-Gilbert (LLG) equation equipped with the optimal switching pulse as an external field. The LLG equation was integrated numerically using the semi-implicit scheme B as described in Ref.~\cite{mentink2010stable}. Particular settings for studying
effects of temperature and material parameter perturbations are described in what follows.

\textit{Effect of thermal fluctuations. }
Interaction of the magnetic systems with the heat bath was simulated by including a stochastic term in the LLG equation. Each simulation had three stages: 1) Initial equilibration at zero applied magnetic field to establish Boltzmann distribution; 2) Switching where the optimal magnetic field is applied (note that thermal fluctuations were also included during the switching stage); 3) Final equilibration at zero applied magnetic field. The duration of the switching stage, i.e. the switching time, was chosen to be $T=2 \tau_0$, while the dimensionless parameter defining the relative strength of the hard-axis anisotropy was $\xi=5$. At the end of the third stage, we inspected the value of $s_z$; we have taken the value $s_z=-0.5$ as the threshold for the successful switching. 

For each value of temperature and damping constant, we repeated simulations $L=1000$ times in order to accumulate the proper statistics. The switching success rate is defined as $f=L_s/L$ where $L_s$ is the number of successful reversals. We find that when the ratio $\Delta E/\Theta>30$, with $\Delta E=K$ [see Eq.~(\ref{energy_biaxial})] being the energy barrier between the stable states and $\Theta$ being thermal energy, the success rate is over 90 \%, see Table~\ref{table1:AppendixB}. For $\Delta E/\Theta \gtrsim 60$, which is a standard requirement to ensure sufficient stability of the magnetic element with respect to thermal fluctuations so as to prevent data loss in magnetic memories~\cite{richter2009density,krounbi2015keynote}, the success rate is close to unity. This result demonstrates that the optimal switching protocol is robust with respect to thermal fluctuations in the technologically relevant regime.

\begin{table}[t!]
 \caption{Magnetization reversal success rate, $f$, for several values of the damping factor $\alpha$, and the ratio $\Delta E/\Theta$, with $\Delta E$ being the energy barrier between the stable states and $\Theta$ being thermal energy.}
    \centering
    \begin{tabular}{p{2.5cm}|p{2.5cm}|p{2.5cm}}
    \hline\hline
   \hfil $\Delta E/\Theta$ & \hfil $\alpha$  &\hfil $f$ (\%)  \\
    \hline
   \hfil  80 & \hfil 0.01 & \hfil 99.9 \\
   \hfil  80 & \hfil 0.1 & \hfil 99.8\\
   \hfil 70 & \hfil 0.01 & \hfil 99.6\\
   \hfil 70 & \hfil 0.1 &  \hfil 99.6\\
   \hfil 50 & \hfil 0.01 & \hfil  98.4\\
   \hfil 50 & \hfil 0.1 & \hfil  98.9\\
   \hfil 30 & \hfil 0.01 &  \hfil 95.3\\
   \hfil 30 & \hfil 0.1 &  \hfil 96.8\\
    \hline\hline
    \end{tabular}
    \label{table1:AppendixB}
\end{table}

\textit{Effect of perturbations in the material parameter values. }
Parameters determining the magnetization dynamics of the monodomain particle include the damping factor $\alpha$, anisotropy parameter $K$, the relative strength of the hard-axis anisotropy $\xi$, and magnetic moment $\mu$. Since $K$ and $\xi$ enter the equation of motion through the parameter $\tau_0$, we only consider perturbations in $\alpha$, $\xi$, and $\tau_0$. In particular, we applied an optimal field pulse derived for a certain value of $\tau_0$, $\xi$, and $\alpha$ to a particle characterized by perturbed parameter values $\tau_0 + \Delta \tau_0$, $\alpha + \Delta \alpha$, and $\xi + \Delta \xi$. The switching time and unperturbed material parameters were chosen to be $T=2\tau_0$, $\alpha=0.1$, $\xi = 5$. Figure~\ref{fig10} shows the results of these calculations. The switching pulse brings the
system over the energy barrier if the strength of the parameter perturbations is not too large.

\bibliographystyle{apsrev4-2}
\bibliography{references}

\begin{thebibliography}{60}%
\makeatletter
\providecommand \@ifxundefined [1]{%
 \@ifx{#1\undefined}
}%
\providecommand \@ifnum [1]{%
 \ifnum #1\expandafter \@firstoftwo
 \else \expandafter \@secondoftwo
 \fi
}%
\providecommand \@ifx [1]{%
 \ifx #1\expandafter \@firstoftwo
 \else \expandafter \@secondoftwo
 \fi
}%
\providecommand \natexlab [1]{#1}%
\providecommand \enquote  [1]{``#1''}%
\providecommand \bibnamefont  [1]{#1}%
\providecommand \bibfnamefont [1]{#1}%
\providecommand \citenamefont [1]{#1}%
\providecommand \href@noop [0]{\@secondoftwo}%
\providecommand \href [0]{\begingroup \@sanitize@url \@href}%
\providecommand \@href[1]{\@@startlink{#1}\@@href}%
\providecommand \@@href[1]{\endgroup#1\@@endlink}%
\providecommand \@sanitize@url [0]{\catcode `\\12\catcode `\$12\catcode
  `\&12\catcode `\#12\catcode `\^12\catcode `\_12\catcode `\%12\relax}%
\providecommand \@@startlink[1]{}%
\providecommand \@@endlink[0]{}%
\providecommand \url  [0]{\begingroup\@sanitize@url \@url }%
\providecommand \@url [1]{\endgroup\@href {#1}{\urlprefix }}%
\providecommand \urlprefix  [0]{URL }%
\providecommand \Eprint [0]{\href }%
\providecommand \doibase [0]{https://doi.org/}%
\providecommand \selectlanguage [0]{\@gobble}%
\providecommand \bibinfo  [0]{\@secondoftwo}%
\providecommand \bibfield  [0]{\@secondoftwo}%
\providecommand \translation [1]{[#1]}%
\providecommand \BibitemOpen [0]{}%
\providecommand \bibitemStop [0]{}%
\providecommand \bibitemNoStop [0]{.\EOS\space}%
\providecommand \EOS [0]{\spacefactor3000\relax}%
\providecommand \BibitemShut  [1]{\csname bibitem#1\endcsname}%
\let\auto@bib@innerbib\@empty
\bibitem [{\citenamefont {Vomir}\ \emph {et~al.}(2005)\citenamefont {Vomir},
  \citenamefont {Andrade}, \citenamefont {Guidoni}, \citenamefont
  {Beaurepaire},\ and\ \citenamefont {Bigot}}]{vomir2005real}%
  \BibitemOpen
  \bibfield  {author} {\bibinfo {author} {\bibfnamefont {M.}~\bibnamefont
  {Vomir}}, \bibinfo {author} {\bibfnamefont {L.~H.~F.}\ \bibnamefont
  {Andrade}}, \bibinfo {author} {\bibfnamefont {L.}~\bibnamefont {Guidoni}},
  \bibinfo {author} {\bibfnamefont {E.}~\bibnamefont {Beaurepaire}},\ and\
  \bibinfo {author} {\bibfnamefont {J.-Y.}\ \bibnamefont {Bigot}},\ }\href
  {https://doi.org/10.1103/PhysRevLett.94.237601} {\bibfield  {journal}
  {\bibinfo  {journal} {Phys. Rev. Lett.}\ }\textbf {\bibinfo {volume} {94}},\
  \bibinfo {pages} {237601} (\bibinfo {year} {2005})}\BibitemShut {NoStop}%
\bibitem [{\citenamefont {Hohlfeld}\ \emph {et~al.}(2001)\citenamefont
  {Hohlfeld}, \citenamefont {Gerrits}, \citenamefont {Bilderbeek},
  \citenamefont {Rasing}, \citenamefont {Awano},\ and\ \citenamefont
  {Ohta}}]{hohlfeld2001fast}%
  \BibitemOpen
  \bibfield  {author} {\bibinfo {author} {\bibfnamefont {J.}~\bibnamefont
  {Hohlfeld}}, \bibinfo {author} {\bibfnamefont {T.}~\bibnamefont {Gerrits}},
  \bibinfo {author} {\bibfnamefont {M.}~\bibnamefont {Bilderbeek}}, \bibinfo
  {author} {\bibfnamefont {T.}~\bibnamefont {Rasing}}, \bibinfo {author}
  {\bibfnamefont {H.}~\bibnamefont {Awano}},\ and\ \bibinfo {author}
  {\bibfnamefont {N.}~\bibnamefont {Ohta}},\ }\href
  {https://doi.org/10.1103/PhysRevB.65.012413} {\bibfield  {journal} {\bibinfo
  {journal} {Phys. Rev. B}\ }\textbf {\bibinfo {volume} {65}},\ \bibinfo
  {pages} {012413} (\bibinfo {year} {2001})}\BibitemShut {NoStop}%
\bibitem [{\citenamefont {Le~Guyader}\ \emph {et~al.}(2012)\citenamefont
  {Le~Guyader}, \citenamefont {El~Moussaoui}, \citenamefont {Buzzi},
  \citenamefont {Chopdekar}, \citenamefont {Heyderman}, \citenamefont
  {Tsukamoto}, \citenamefont {Itoh}, \citenamefont {Kirilyuk}, \citenamefont
  {Rasing}, \citenamefont {Kimel},\ and\ \citenamefont
  {Nolting}}]{le2012demonstration}%
  \BibitemOpen
  \bibfield  {author} {\bibinfo {author} {\bibfnamefont {L.}~\bibnamefont
  {Le~Guyader}}, \bibinfo {author} {\bibfnamefont {S.}~\bibnamefont
  {El~Moussaoui}}, \bibinfo {author} {\bibfnamefont {M.}~\bibnamefont {Buzzi}},
  \bibinfo {author} {\bibfnamefont {R.~V.}\ \bibnamefont {Chopdekar}}, \bibinfo
  {author} {\bibfnamefont {L.~J.}\ \bibnamefont {Heyderman}}, \bibinfo {author}
  {\bibfnamefont {A.}~\bibnamefont {Tsukamoto}}, \bibinfo {author}
  {\bibfnamefont {A.}~\bibnamefont {Itoh}}, \bibinfo {author} {\bibfnamefont
  {A.}~\bibnamefont {Kirilyuk}}, \bibinfo {author} {\bibfnamefont
  {T.}~\bibnamefont {Rasing}}, \bibinfo {author} {\bibfnamefont {A.~V.}\
  \bibnamefont {Kimel}},\ and\ \bibinfo {author} {\bibfnamefont
  {F.}~\bibnamefont {Nolting}},\ }\href {https://doi.org/10.1063/1.4733965}
  {\bibfield  {journal} {\bibinfo  {journal} {Appl. Phys. Lett.}\ }\textbf
  {\bibinfo {volume} {101}},\ \bibinfo {pages} {022410} (\bibinfo {year}
  {2012})}\BibitemShut {NoStop}%
\bibitem [{\citenamefont {Myers}\ \emph {et~al.}(1999)\citenamefont {Myers},
  \citenamefont {Ralph}, \citenamefont {Katine}, \citenamefont {Louie},\ and\
  \citenamefont {Buhrman}}]{myers1999current}%
  \BibitemOpen
  \bibfield  {author} {\bibinfo {author} {\bibfnamefont {E.~B.}\ \bibnamefont
  {Myers}}, \bibinfo {author} {\bibfnamefont {D.~C.}\ \bibnamefont {Ralph}},
  \bibinfo {author} {\bibfnamefont {J.~A.}\ \bibnamefont {Katine}}, \bibinfo
  {author} {\bibfnamefont {R.~N.}\ \bibnamefont {Louie}},\ and\ \bibinfo
  {author} {\bibfnamefont {R.~A.}\ \bibnamefont {Buhrman}},\ }\href
  {https://doi.org/10.1126/science.285.5429.867} {\bibfield  {journal}
  {\bibinfo  {journal} {Science}\ }\textbf {\bibinfo {volume} {285}},\ \bibinfo
  {pages} {867} (\bibinfo {year} {1999})}\BibitemShut {NoStop}%
\bibitem [{\citenamefont {Katine}\ \emph {et~al.}(2000)\citenamefont {Katine},
  \citenamefont {Albert}, \citenamefont {Buhrman}, \citenamefont {Myers},\ and\
  \citenamefont {Ralph}}]{katine2000current}%
  \BibitemOpen
  \bibfield  {author} {\bibinfo {author} {\bibfnamefont {J.~A.}\ \bibnamefont
  {Katine}}, \bibinfo {author} {\bibfnamefont {F.~J.}\ \bibnamefont {Albert}},
  \bibinfo {author} {\bibfnamefont {R.~A.}\ \bibnamefont {Buhrman}}, \bibinfo
  {author} {\bibfnamefont {E.~B.}\ \bibnamefont {Myers}},\ and\ \bibinfo
  {author} {\bibfnamefont {D.~C.}\ \bibnamefont {Ralph}},\ }\href
  {https://doi.org/10.1103/PhysRevLett.84.3149} {\bibfield  {journal} {\bibinfo
   {journal} {Phys. Rev. Lett.}\ }\textbf {\bibinfo {volume} {84}},\ \bibinfo
  {pages} {3149} (\bibinfo {year} {2000})}\BibitemShut {NoStop}%
\bibitem [{\citenamefont {Sun}\ and\ \citenamefont {Wang}(2005)}]{sun2005fast}%
  \BibitemOpen
  \bibfield  {author} {\bibinfo {author} {\bibfnamefont {Z.~Z.}\ \bibnamefont
  {Sun}}\ and\ \bibinfo {author} {\bibfnamefont {X.~R.}\ \bibnamefont {Wang}},\
  }\href {https://doi.org/10.1103/PhysRevB.71.174430} {\bibfield  {journal}
  {\bibinfo  {journal} {Phys. Rev. B}\ }\textbf {\bibinfo {volume} {71}},\
  \bibinfo {pages} {174430} (\bibinfo {year} {2005})}\BibitemShut {NoStop}%
\bibitem [{\citenamefont {Acremann}\ \emph {et~al.}(2000)\citenamefont
  {Acremann}, \citenamefont {Back}, \citenamefont {Buess}, \citenamefont
  {Portmann}, \citenamefont {Vaterlaus}, \citenamefont {Pescia},\ and\
  \citenamefont {Melchior}}]{acremann2000imaging}%
  \BibitemOpen
  \bibfield  {author} {\bibinfo {author} {\bibfnamefont {Y.}~\bibnamefont
  {Acremann}}, \bibinfo {author} {\bibfnamefont {C.~H.}\ \bibnamefont {Back}},
  \bibinfo {author} {\bibfnamefont {M.}~\bibnamefont {Buess}}, \bibinfo
  {author} {\bibfnamefont {O.}~\bibnamefont {Portmann}}, \bibinfo {author}
  {\bibfnamefont {A.}~\bibnamefont {Vaterlaus}}, \bibinfo {author}
  {\bibfnamefont {D.}~\bibnamefont {Pescia}},\ and\ \bibinfo {author}
  {\bibfnamefont {H.}~\bibnamefont {Melchior}},\ }\href
  {https://doi.org/10.1126/science.290.5491.492} {\bibfield  {journal}
  {\bibinfo  {journal} {Science}\ }\textbf {\bibinfo {volume} {290}},\ \bibinfo
  {pages} {492} (\bibinfo {year} {2000})}\BibitemShut {NoStop}%
\bibitem [{\citenamefont {Hiebert}\ \emph {et~al.}(1997)\citenamefont
  {Hiebert}, \citenamefont {Stankiewicz},\ and\ \citenamefont
  {Freeman}}]{hiebert1997direct}%
  \BibitemOpen
  \bibfield  {author} {\bibinfo {author} {\bibfnamefont {W.~K.}\ \bibnamefont
  {Hiebert}}, \bibinfo {author} {\bibfnamefont {A.}~\bibnamefont
  {Stankiewicz}},\ and\ \bibinfo {author} {\bibfnamefont {M.~R.}\ \bibnamefont
  {Freeman}},\ }\href {https://doi.org/10.1103/PhysRevLett.79.1134} {\bibfield
  {journal} {\bibinfo  {journal} {Phys. Rev. Lett.}\ }\textbf {\bibinfo
  {volume} {79}},\ \bibinfo {pages} {1134} (\bibinfo {year}
  {1997})}\BibitemShut {NoStop}%
\bibitem [{\citenamefont {Xiao}\ \emph {et~al.}(2006)\citenamefont {Xiao},
  \citenamefont {Tsoi},\ and\ \citenamefont {Niu}}]{xiao2006minimal}%
  \BibitemOpen
  \bibfield  {author} {\bibinfo {author} {\bibfnamefont {D.}~\bibnamefont
  {Xiao}}, \bibinfo {author} {\bibfnamefont {M.}~\bibnamefont {Tsoi}},\ and\
  \bibinfo {author} {\bibfnamefont {Q.}~\bibnamefont {Niu}},\ }\href
  {https://doi.org/10.1063/1.2161421} {\bibfield  {journal} {\bibinfo
  {journal} {J. Appl. Phys.}\ }\textbf {\bibinfo {volume} {99}},\ \bibinfo
  {pages} {013903} (\bibinfo {year} {2006})}\BibitemShut {NoStop}%
\bibitem [{\citenamefont {Yang}\ \emph {et~al.}(2017)\citenamefont {Yang},
  \citenamefont {Wilson}, \citenamefont {Gorchon}, \citenamefont {Lambert},
  \citenamefont {Salahuddin},\ and\ \citenamefont {Bokor}}]{yang2017ultrafast}%
  \BibitemOpen
  \bibfield  {author} {\bibinfo {author} {\bibfnamefont {Y.}~\bibnamefont
  {Yang}}, \bibinfo {author} {\bibfnamefont {R.~B.}\ \bibnamefont {Wilson}},
  \bibinfo {author} {\bibfnamefont {J.}~\bibnamefont {Gorchon}}, \bibinfo
  {author} {\bibfnamefont {C.-H.}\ \bibnamefont {Lambert}}, \bibinfo {author}
  {\bibfnamefont {S.}~\bibnamefont {Salahuddin}},\ and\ \bibinfo {author}
  {\bibfnamefont {J.}~\bibnamefont {Bokor}},\ }\href
  {https://doi.org/10.1126/sciadv.1603117} {\bibfield  {journal} {\bibinfo
  {journal} {Sci. Adv.}\ }\textbf {\bibinfo {volume} {3}},\ \bibinfo {pages}
  {e1603117} (\bibinfo {year} {2017})}\BibitemShut {NoStop}%
\bibitem [{\citenamefont {Thirion}\ \emph {et~al.}(2003)\citenamefont
  {Thirion}, \citenamefont {Wernsdorfer},\ and\ \citenamefont
  {Mailly}}]{thirion2003switching}%
  \BibitemOpen
  \bibfield  {author} {\bibinfo {author} {\bibfnamefont {C.}~\bibnamefont
  {Thirion}}, \bibinfo {author} {\bibfnamefont {W.}~\bibnamefont
  {Wernsdorfer}},\ and\ \bibinfo {author} {\bibfnamefont {D.}~\bibnamefont
  {Mailly}},\ }\href {https://doi.org/10.1038/nmat946} {\bibfield  {journal}
  {\bibinfo  {journal} {Nat. Mater.}\ }\textbf {\bibinfo {volume} {2}},\
  \bibinfo {pages} {524} (\bibinfo {year} {2003})}\BibitemShut {NoStop}%
\bibitem [{\citenamefont {Sun}\ and\ \citenamefont
  {Wang}(2006)}]{sun2006magnetization}%
  \BibitemOpen
  \bibfield  {author} {\bibinfo {author} {\bibfnamefont {Z.~Z.}\ \bibnamefont
  {Sun}}\ and\ \bibinfo {author} {\bibfnamefont {X.~R.}\ \bibnamefont {Wang}},\
  }\href {https://doi.org/10.1103/PhysRevB.74.132401} {\bibfield  {journal}
  {\bibinfo  {journal} {Phys. Rev. B}\ }\textbf {\bibinfo {volume} {74}},\
  \bibinfo {pages} {132401} (\bibinfo {year} {2006})}\BibitemShut {NoStop}%
\bibitem [{\citenamefont {Cai}\ \emph {et~al.}(2013)\citenamefont {Cai},
  \citenamefont {Garanin},\ and\ \citenamefont {Chudnovsky}}]{cai2013reversal}%
  \BibitemOpen
  \bibfield  {author} {\bibinfo {author} {\bibfnamefont {L.}~\bibnamefont
  {Cai}}, \bibinfo {author} {\bibfnamefont {D.~A.}\ \bibnamefont {Garanin}},\
  and\ \bibinfo {author} {\bibfnamefont {E.~M.}\ \bibnamefont {Chudnovsky}},\
  }\href {https://doi.org/10.1103/PhysRevB.87.024418} {\bibfield  {journal}
  {\bibinfo  {journal} {Phys. Rev. B}\ }\textbf {\bibinfo {volume} {87}},\
  \bibinfo {pages} {024418} (\bibinfo {year} {2013})}\BibitemShut {NoStop}%
\bibitem [{\citenamefont {Bandyopadhyay}\ \emph {et~al.}(2021)\citenamefont
  {Bandyopadhyay}, \citenamefont {Atulasimha},\ and\ \citenamefont
  {Barman}}]{bandyopadhyay2021magnetic}%
  \BibitemOpen
  \bibfield  {author} {\bibinfo {author} {\bibfnamefont {S.}~\bibnamefont
  {Bandyopadhyay}}, \bibinfo {author} {\bibfnamefont {J.}~\bibnamefont
  {Atulasimha}},\ and\ \bibinfo {author} {\bibfnamefont {A.}~\bibnamefont
  {Barman}},\ }\href {https://doi.org/10.1063/5.0062993} {\bibfield  {journal}
  {\bibinfo  {journal} {Appl. Phys. Rev.}\ }\textbf {\bibinfo {volume} {8}},\
  \bibinfo {pages} {041323} (\bibinfo {year} {2021})}\BibitemShut {NoStop}%
\bibitem [{\citenamefont {Pushp}\ \emph {et~al.}(2015)\citenamefont {Pushp},
  \citenamefont {Phung}, \citenamefont {Rettner}, \citenamefont {Hughes},
  \citenamefont {Yang},\ and\ \citenamefont {Parkin}}]{Pushp2015}%
  \BibitemOpen
  \bibfield  {author} {\bibinfo {author} {\bibfnamefont {A.}~\bibnamefont
  {Pushp}}, \bibinfo {author} {\bibfnamefont {T.}~\bibnamefont {Phung}},
  \bibinfo {author} {\bibfnamefont {C.}~\bibnamefont {Rettner}}, \bibinfo
  {author} {\bibfnamefont {B.~P.}\ \bibnamefont {Hughes}}, \bibinfo {author}
  {\bibfnamefont {S.-H.}\ \bibnamefont {Yang}},\ and\ \bibinfo {author}
  {\bibfnamefont {S.~S.~P.}\ \bibnamefont {Parkin}},\ }\href
  {https://doi.org/10.1073/pnas.1507084112} {\bibfield  {journal} {\bibinfo
  {journal} {Proc. Natl. Acad. Sci. U.S.A.}\ }\textbf {\bibinfo {volume}
  {112}},\ \bibinfo {pages} {6585} (\bibinfo {year} {2015})}\BibitemShut
  {NoStop}%
\bibitem [{\citenamefont {Michel}\ \emph {et~al.}(2017)\citenamefont {Michel},
  \citenamefont {Niemann}, \citenamefont {Boehnert}, \citenamefont {Martens},
  \citenamefont {Moreno}, \citenamefont {Goerlitz}, \citenamefont {Zierold},
  \citenamefont {Reith}, \citenamefont {Vega}, \citenamefont {Prida},
  \citenamefont {Thomas}, \citenamefont {Gooth},\ and\ \citenamefont
  {Nielsch}}]{Michel_2017}%
  \BibitemOpen
  \bibfield  {author} {\bibinfo {author} {\bibfnamefont {A.-K.}\ \bibnamefont
  {Michel}}, \bibinfo {author} {\bibfnamefont {A.~C.}\ \bibnamefont {Niemann}},
  \bibinfo {author} {\bibfnamefont {T.}~\bibnamefont {Boehnert}}, \bibinfo
  {author} {\bibfnamefont {S.}~\bibnamefont {Martens}}, \bibinfo {author}
  {\bibfnamefont {J.~M.~M.}\ \bibnamefont {Moreno}}, \bibinfo {author}
  {\bibfnamefont {D.}~\bibnamefont {Goerlitz}}, \bibinfo {author}
  {\bibfnamefont {R.}~\bibnamefont {Zierold}}, \bibinfo {author} {\bibfnamefont
  {H.}~\bibnamefont {Reith}}, \bibinfo {author} {\bibfnamefont
  {V.}~\bibnamefont {Vega}}, \bibinfo {author} {\bibfnamefont {V.~M.}\
  \bibnamefont {Prida}}, \bibinfo {author} {\bibfnamefont {A.}~\bibnamefont
  {Thomas}}, \bibinfo {author} {\bibfnamefont {J.}~\bibnamefont {Gooth}},\ and\
  \bibinfo {author} {\bibfnamefont {K.}~\bibnamefont {Nielsch}},\ }\href
  {https://doi.org/10.1088/1361-6463/aa9444} {\bibfield  {journal} {\bibinfo
  {journal} {J. Phys. D}\ }\textbf {\bibinfo {volume} {50}},\ \bibinfo {pages}
  {494007} (\bibinfo {year} {2017})}\BibitemShut {NoStop}%
\bibitem [{\citenamefont {Bertotti}\ \emph
  {et~al.}(2003{\natexlab{a}})\citenamefont {Bertotti}, \citenamefont
  {Mayergoyz}, \citenamefont {Serpico},\ and\ \citenamefont
  {Dimian}}]{bertotti2003comparison}%
  \BibitemOpen
  \bibfield  {author} {\bibinfo {author} {\bibfnamefont {G.}~\bibnamefont
  {Bertotti}}, \bibinfo {author} {\bibfnamefont {I.}~\bibnamefont {Mayergoyz}},
  \bibinfo {author} {\bibfnamefont {C.}~\bibnamefont {Serpico}},\ and\ \bibinfo
  {author} {\bibfnamefont {M.}~\bibnamefont {Dimian}},\ }\href
  {https://doi.org/10.1063/1.1557275} {\bibfield  {journal} {\bibinfo
  {journal} {J. Appl. Phys.}\ }\textbf {\bibinfo {volume} {93}},\ \bibinfo
  {pages} {6811} (\bibinfo {year} {2003}{\natexlab{a}})}\BibitemShut {NoStop}%
\bibitem [{\citenamefont {Mallinson}(2000)}]{mallinson2000damped}%
  \BibitemOpen
  \bibfield  {author} {\bibinfo {author} {\bibfnamefont {J.}~\bibnamefont
  {Mallinson}},\ }\href {https://doi.org/10.1109/20.875251} {\bibfield
  {journal} {\bibinfo  {journal} {IEEE Trans. Magn.}\ }\textbf {\bibinfo
  {volume} {36}},\ \bibinfo {pages} {1976} (\bibinfo {year}
  {2000})}\BibitemShut {NoStop}%
\bibitem [{\citenamefont {Suess}\ \emph {et~al.}(2009)\citenamefont {Suess},
  \citenamefont {Lee}, \citenamefont {Fidler},\ and\ \citenamefont
  {Schrefl}}]{suess2009exchange}%
  \BibitemOpen
  \bibfield  {author} {\bibinfo {author} {\bibfnamefont {D.}~\bibnamefont
  {Suess}}, \bibinfo {author} {\bibfnamefont {J.}~\bibnamefont {Lee}}, \bibinfo
  {author} {\bibfnamefont {J.}~\bibnamefont {Fidler}},\ and\ \bibinfo {author}
  {\bibfnamefont {T.}~\bibnamefont {Schrefl}},\ }\href
  {https://doi.org/https://doi.org/10.1016/j.jmmm.2008.06.041} {\bibfield
  {journal} {\bibinfo  {journal} {J. Magn. Magn. Mater.}\ }\textbf {\bibinfo
  {volume} {321}},\ \bibinfo {pages} {545} (\bibinfo {year}
  {2009})}\BibitemShut {NoStop}%
\bibitem [{\citenamefont {Back}\ \emph {et~al.}(1998)\citenamefont {Back},
  \citenamefont {Weller}, \citenamefont {Heidmann}, \citenamefont {Mauri},
  \citenamefont {Guarisco}, \citenamefont {Garwin},\ and\ \citenamefont
  {Siegmann}}]{back1998magnetization}%
  \BibitemOpen
  \bibfield  {author} {\bibinfo {author} {\bibfnamefont {C.~H.}\ \bibnamefont
  {Back}}, \bibinfo {author} {\bibfnamefont {D.}~\bibnamefont {Weller}},
  \bibinfo {author} {\bibfnamefont {J.}~\bibnamefont {Heidmann}}, \bibinfo
  {author} {\bibfnamefont {D.}~\bibnamefont {Mauri}}, \bibinfo {author}
  {\bibfnamefont {D.}~\bibnamefont {Guarisco}}, \bibinfo {author}
  {\bibfnamefont {E.~L.}\ \bibnamefont {Garwin}},\ and\ \bibinfo {author}
  {\bibfnamefont {H.~C.}\ \bibnamefont {Siegmann}},\ }\href
  {https://doi.org/10.1103/PhysRevLett.81.3251} {\bibfield  {journal} {\bibinfo
   {journal} {Phys. Rev. Lett.}\ }\textbf {\bibinfo {volume} {81}},\ \bibinfo
  {pages} {3251} (\bibinfo {year} {1998})}\BibitemShut {NoStop}%
\bibitem [{\citenamefont {Bertotti}\ \emph
  {et~al.}(2003{\natexlab{b}})\citenamefont {Bertotti}, \citenamefont
  {Mayergoyz},\ and\ \citenamefont {Serpico}}]{bertotti2003critical}%
  \BibitemOpen
  \bibfield  {author} {\bibinfo {author} {\bibfnamefont {G.}~\bibnamefont
  {Bertotti}}, \bibinfo {author} {\bibfnamefont {I.}~\bibnamefont
  {Mayergoyz}},\ and\ \bibinfo {author} {\bibfnamefont {C.}~\bibnamefont
  {Serpico}},\ }\href {https://doi.org/10.1109/TMAG.2003.816454} {\bibfield
  {journal} {\bibinfo  {journal} {IEEE Trans. Magn.}\ }\textbf {\bibinfo
  {volume} {39}},\ \bibinfo {pages} {2504} (\bibinfo {year}
  {2003}{\natexlab{b}})}\BibitemShut {NoStop}%
\bibitem [{\citenamefont {Bauer}\ \emph {et~al.}(2000)\citenamefont {Bauer},
  \citenamefont {Lopusnik}, \citenamefont {Fassbender},\ and\ \citenamefont
  {Hillebrands}}]{bauer2000suppression}%
  \BibitemOpen
  \bibfield  {author} {\bibinfo {author} {\bibfnamefont {M.}~\bibnamefont
  {Bauer}}, \bibinfo {author} {\bibfnamefont {R.}~\bibnamefont {Lopusnik}},
  \bibinfo {author} {\bibfnamefont {J.}~\bibnamefont {Fassbender}},\ and\
  \bibinfo {author} {\bibfnamefont {B.}~\bibnamefont {Hillebrands}},\ }\href
  {https://doi.org/10.1063/1.126466} {\bibfield  {journal} {\bibinfo  {journal}
  {Appl. Phys. Lett.}\ }\textbf {\bibinfo {volume} {76}},\ \bibinfo {pages}
  {2758} (\bibinfo {year} {2000})}\BibitemShut {NoStop}%
\bibitem [{\citenamefont {Gerrits}\ \emph {et~al.}(2002)\citenamefont
  {Gerrits}, \citenamefont {van~den Berg}, \citenamefont {Hohlfeld},
  \citenamefont {B{\"a}r},\ and\ \citenamefont
  {Rasing}}]{gerrits2002ultrafast}%
  \BibitemOpen
  \bibfield  {author} {\bibinfo {author} {\bibfnamefont {T.}~\bibnamefont
  {Gerrits}}, \bibinfo {author} {\bibfnamefont {H.~A.~M.}\ \bibnamefont
  {van~den Berg}}, \bibinfo {author} {\bibfnamefont {J.}~\bibnamefont
  {Hohlfeld}}, \bibinfo {author} {\bibfnamefont {L.}~\bibnamefont {B{\"a}r}},\
  and\ \bibinfo {author} {\bibfnamefont {T.}~\bibnamefont {Rasing}},\ }\href
  {https://doi.org/10.1038/nature00905} {\bibfield  {journal} {\bibinfo
  {journal} {Nature}\ }\textbf {\bibinfo {volume} {418}},\ \bibinfo {pages}
  {509} (\bibinfo {year} {2002})}\BibitemShut {NoStop}%
\bibitem [{\citenamefont {Schumacher}\ \emph {et~al.}(2003)\citenamefont
  {Schumacher}, \citenamefont {Chappert}, \citenamefont {Crozat}, \citenamefont
  {Sousa}, \citenamefont {Freitas}, \citenamefont {Miltat}, \citenamefont
  {Fassbender},\ and\ \citenamefont {Hillebrands}}]{schumacher2003phase}%
  \BibitemOpen
  \bibfield  {author} {\bibinfo {author} {\bibfnamefont {H.~W.}\ \bibnamefont
  {Schumacher}}, \bibinfo {author} {\bibfnamefont {C.}~\bibnamefont
  {Chappert}}, \bibinfo {author} {\bibfnamefont {P.}~\bibnamefont {Crozat}},
  \bibinfo {author} {\bibfnamefont {R.~C.}\ \bibnamefont {Sousa}}, \bibinfo
  {author} {\bibfnamefont {P.~P.}\ \bibnamefont {Freitas}}, \bibinfo {author}
  {\bibfnamefont {J.}~\bibnamefont {Miltat}}, \bibinfo {author} {\bibfnamefont
  {J.}~\bibnamefont {Fassbender}},\ and\ \bibinfo {author} {\bibfnamefont
  {B.}~\bibnamefont {Hillebrands}},\ }\href
  {https://doi.org/10.1103/PhysRevLett.90.017201} {\bibfield  {journal}
  {\bibinfo  {journal} {Phys. Rev. Lett.}\ }\textbf {\bibinfo {volume} {90}},\
  \bibinfo {pages} {017201} (\bibinfo {year} {2003})}\BibitemShut {NoStop}%
\bibitem [{\citenamefont {Rivkin}\ and\ \citenamefont
  {Ketterson}(2006)}]{rivkin2006magnetization}%
  \BibitemOpen
  \bibfield  {author} {\bibinfo {author} {\bibfnamefont {K.}~\bibnamefont
  {Rivkin}}\ and\ \bibinfo {author} {\bibfnamefont {J.~B.}\ \bibnamefont
  {Ketterson}},\ }\href {https://doi.org/10.1063/1.2405855} {\bibfield
  {journal} {\bibinfo  {journal} {Appl. Phys. Lett.}\ }\textbf {\bibinfo
  {volume} {89}},\ \bibinfo {pages} {252507} (\bibinfo {year}
  {2006})}\BibitemShut {NoStop}%
\bibitem [{\citenamefont {Pontryagin}\ \emph {et~al.}(1962)\citenamefont
  {Pontryagin}, \citenamefont {Boltyanskii}, \citenamefont {Gamkrelidze},\ and\
  \citenamefont {Mishchenko}}]{pontryagin2018mathematical}%
  \BibitemOpen
  \bibfield  {author} {\bibinfo {author} {\bibfnamefont {L.~S.}\ \bibnamefont
  {Pontryagin}}, \bibinfo {author} {\bibfnamefont {V.~G.}\ \bibnamefont
  {Boltyanskii}}, \bibinfo {author} {\bibfnamefont {R.~V.}\ \bibnamefont
  {Gamkrelidze}},\ and\ \bibinfo {author} {\bibfnamefont {E.~F.}\ \bibnamefont
  {Mishchenko}},\ }\href@noop {} {\emph {\bibinfo {title} {The Mathematical
  Theory of Optimal Processes}}}\ (\bibinfo  {publisher} {Interscience},\
  \bibinfo {address} {New York},\ \bibinfo {year} {1962})\BibitemShut {NoStop}%
\bibitem [{\citenamefont {Barros}\ \emph {et~al.}(2011)\citenamefont {Barros},
  \citenamefont {Rassam}, \citenamefont {Jirari},\ and\ \citenamefont
  {Kachkachi}}]{barros2011optimal}%
  \BibitemOpen
  \bibfield  {author} {\bibinfo {author} {\bibfnamefont {N.}~\bibnamefont
  {Barros}}, \bibinfo {author} {\bibfnamefont {M.}~\bibnamefont {Rassam}},
  \bibinfo {author} {\bibfnamefont {H.}~\bibnamefont {Jirari}},\ and\ \bibinfo
  {author} {\bibfnamefont {H.}~\bibnamefont {Kachkachi}},\ }\href
  {https://doi.org/10.1103/PhysRevB.83.144418} {\bibfield  {journal} {\bibinfo
  {journal} {Phys. Rev. B}\ }\textbf {\bibinfo {volume} {83}},\ \bibinfo
  {pages} {144418} (\bibinfo {year} {2011})}\BibitemShut {NoStop}%
\bibitem [{\citenamefont {Barros}\ \emph {et~al.}(2013)\citenamefont {Barros},
  \citenamefont {Rassam},\ and\ \citenamefont
  {Kachkachi}}]{barros2013microwave}%
  \BibitemOpen
  \bibfield  {author} {\bibinfo {author} {\bibfnamefont {N.}~\bibnamefont
  {Barros}}, \bibinfo {author} {\bibfnamefont {H.}~\bibnamefont {Rassam}},\
  and\ \bibinfo {author} {\bibfnamefont {H.}~\bibnamefont {Kachkachi}},\ }\href
  {https://doi.org/10.1103/PhysRevB.88.014421} {\bibfield  {journal} {\bibinfo
  {journal} {Phys. Rev. B}\ }\textbf {\bibinfo {volume} {88}},\ \bibinfo
  {pages} {014421} (\bibinfo {year} {2013})}\BibitemShut {NoStop}%
\bibitem [{\citenamefont {Kwiatkowski}\ \emph {et~al.}(2021)\citenamefont
  {Kwiatkowski}, \citenamefont {Badarneh}, \citenamefont {Berkov},\ and\
  \citenamefont {Bessarab}}]{kwiatkowski2021optimal}%
  \BibitemOpen
  \bibfield  {author} {\bibinfo {author} {\bibfnamefont {G.~J.}\ \bibnamefont
  {Kwiatkowski}}, \bibinfo {author} {\bibfnamefont {M.~H.~A.}\ \bibnamefont
  {Badarneh}}, \bibinfo {author} {\bibfnamefont {D.~V.}\ \bibnamefont
  {Berkov}},\ and\ \bibinfo {author} {\bibfnamefont {P.~F.}\ \bibnamefont
  {Bessarab}},\ }\href {https://doi.org/10.1103/PhysRevLett.126.177206}
  {\bibfield  {journal} {\bibinfo  {journal} {Phys. Rev. Lett.}\ }\textbf
  {\bibinfo {volume} {126}},\ \bibinfo {pages} {177206} (\bibinfo {year}
  {2021})}\BibitemShut {NoStop}%
\bibitem [{\citenamefont {Back}\ \emph {et~al.}(1999)\citenamefont {Back},
  \citenamefont {Allenspach}, \citenamefont {Weber}, \citenamefont {Parkin},
  \citenamefont {Weller}, \citenamefont {Garwin},\ and\ \citenamefont
  {Siegmann}}]{back1999minimum}%
  \BibitemOpen
  \bibfield  {author} {\bibinfo {author} {\bibfnamefont {C.~H.}\ \bibnamefont
  {Back}}, \bibinfo {author} {\bibfnamefont {R.}~\bibnamefont {Allenspach}},
  \bibinfo {author} {\bibfnamefont {W.}~\bibnamefont {Weber}}, \bibinfo
  {author} {\bibfnamefont {S.~S.~P.}\ \bibnamefont {Parkin}}, \bibinfo {author}
  {\bibfnamefont {D.}~\bibnamefont {Weller}}, \bibinfo {author} {\bibfnamefont
  {E.~L.}\ \bibnamefont {Garwin}},\ and\ \bibinfo {author} {\bibfnamefont
  {H.~C.}\ \bibnamefont {Siegmann}},\ }\href
  {https://doi.org/10.1126/science.285.5429.864} {\bibfield  {journal}
  {\bibinfo  {journal} {Science}\ }\textbf {\bibinfo {volume} {285}},\ \bibinfo
  {pages} {864} (\bibinfo {year} {1999})}\BibitemShut {NoStop}%
\bibitem [{\citenamefont {Etz}\ \emph {et~al.}(2012)\citenamefont {Etz},
  \citenamefont {Costa}, \citenamefont {Eriksson},\ and\ \citenamefont
  {Bergman}}]{etz2012accelerating}%
  \BibitemOpen
  \bibfield  {author} {\bibinfo {author} {\bibfnamefont {C.}~\bibnamefont
  {Etz}}, \bibinfo {author} {\bibfnamefont {M.}~\bibnamefont {Costa}}, \bibinfo
  {author} {\bibfnamefont {O.}~\bibnamefont {Eriksson}},\ and\ \bibinfo
  {author} {\bibfnamefont {A.}~\bibnamefont {Bergman}},\ }\href
  {https://doi.org/10.1103/PhysRevB.86.224401} {\bibfield  {journal} {\bibinfo
  {journal} {Phys. Rev. B}\ }\textbf {\bibinfo {volume} {86}},\ \bibinfo
  {pages} {224401} (\bibinfo {year} {2012})}\BibitemShut {NoStop}%
\bibitem [{\citenamefont {Osborn}(1945)}]{osborn1945demagnetizing}%
  \BibitemOpen
  \bibfield  {author} {\bibinfo {author} {\bibfnamefont {J.~A.}\ \bibnamefont
  {Osborn}},\ }\href {https://doi.org/10.1103/PhysRev.67.351} {\bibfield
  {journal} {\bibinfo  {journal} {Phys. Rev.}\ }\textbf {\bibinfo {volume}
  {67}},\ \bibinfo {pages} {351} (\bibinfo {year} {1945})}\BibitemShut
  {NoStop}%
\bibitem [{\citenamefont {Chun}\ \emph {et~al.}(2013)\citenamefont {Chun},
  \citenamefont {Zhao}, \citenamefont {Harms}, \citenamefont {Kim},
  \citenamefont {Wang},\ and\ \citenamefont {Kim}}]{chun2012scaling}%
  \BibitemOpen
  \bibfield  {author} {\bibinfo {author} {\bibfnamefont {K.~C.}\ \bibnamefont
  {Chun}}, \bibinfo {author} {\bibfnamefont {H.}~\bibnamefont {Zhao}}, \bibinfo
  {author} {\bibfnamefont {J.~D.}\ \bibnamefont {Harms}}, \bibinfo {author}
  {\bibfnamefont {T.-H.}\ \bibnamefont {Kim}}, \bibinfo {author} {\bibfnamefont
  {J.-P.}\ \bibnamefont {Wang}},\ and\ \bibinfo {author} {\bibfnamefont
  {C.~H.}\ \bibnamefont {Kim}},\ }\href
  {https://doi.org/10.1109/JSSC.2012.2224256} {\bibfield  {journal} {\bibinfo
  {journal} {IEEE J. Solid-State Circuits}\ }\textbf {\bibinfo {volume} {48}},\
  \bibinfo {pages} {598} (\bibinfo {year} {2013})}\BibitemShut {NoStop}%
\bibitem [{\citenamefont {Nisoli}\ \emph {et~al.}(2013)\citenamefont {Nisoli},
  \citenamefont {Moessner},\ and\ \citenamefont
  {Schiffer}}]{nisoli2013colloquium}%
  \BibitemOpen
  \bibfield  {author} {\bibinfo {author} {\bibfnamefont {C.}~\bibnamefont
  {Nisoli}}, \bibinfo {author} {\bibfnamefont {R.}~\bibnamefont {Moessner}},\
  and\ \bibinfo {author} {\bibfnamefont {P.}~\bibnamefont {Schiffer}},\ }\href
  {https://doi.org/10.1103/RevModPhys.85.1473} {\bibfield  {journal} {\bibinfo
  {journal} {Rev. Mod. Phys.}\ }\textbf {\bibinfo {volume} {85}},\ \bibinfo
  {pages} {1473} (\bibinfo {year} {2013})}\BibitemShut {NoStop}%
\bibitem [{\citenamefont {Wysin}\ \emph {et~al.}(2012)\citenamefont {Wysin},
  \citenamefont {Moura-Melo}, \citenamefont {Mól},\ and\ \citenamefont
  {Pereira}}]{wysin2012magnetic}%
  \BibitemOpen
  \bibfield  {author} {\bibinfo {author} {\bibfnamefont {G.~M.}\ \bibnamefont
  {Wysin}}, \bibinfo {author} {\bibfnamefont {W.~A.}\ \bibnamefont
  {Moura-Melo}}, \bibinfo {author} {\bibfnamefont {L.~A.~S.}\ \bibnamefont
  {Mól}},\ and\ \bibinfo {author} {\bibfnamefont {A.~R.}\ \bibnamefont
  {Pereira}},\ }\href {https://doi.org/10.1088/0953-8984/24/29/296001}
  {\bibfield  {journal} {\bibinfo  {journal} {J. Phys. Condens. Matter}\
  }\textbf {\bibinfo {volume} {24}},\ \bibinfo {pages} {296001} (\bibinfo
  {year} {2012})}\BibitemShut {NoStop}%
\bibitem [{\citenamefont {Badarneh}\ \emph {et~al.}(2020)\citenamefont
  {Badarneh}, \citenamefont {Kwiatkowski},\ and\ \citenamefont
  {Bessarab}}]{badarneh2020mechanisms}%
  \BibitemOpen
  \bibfield  {author} {\bibinfo {author} {\bibfnamefont {M.~H.~A.}\
  \bibnamefont {Badarneh}}, \bibinfo {author} {\bibfnamefont {G.~J.}\
  \bibnamefont {Kwiatkowski}},\ and\ \bibinfo {author} {\bibfnamefont {P.~F.}\
  \bibnamefont {Bessarab}},\ }\href
  {https://doi.org/10.17586/2220-8054-2020-11-3-294-300} {\bibfield  {journal}
  {\bibinfo  {journal} {Nanosystems: Physics, Chemistry, Mathematics}\ }\textbf
  {\bibinfo {volume} {11}},\ \bibinfo {pages} {294} (\bibinfo {year}
  {2020})}\BibitemShut {NoStop}%
\bibitem [{\citenamefont {Berkov}(2007)}]{berkov2007magnetization}%
  \BibitemOpen
  \bibfield  {author} {\bibinfo {author} {\bibfnamefont {D.~V.}\ \bibnamefont
  {Berkov}},\ }\href@noop {} {\emph {\bibinfo {title} {Handbook of Magnetism
  and Advanced Magnetic Materials}}},\ edited by\ \bibinfo {editor}
  {\bibfnamefont {H.}~\bibnamefont {Kronm\"uller}}\ and\ \bibinfo {editor}
  {\bibfnamefont {S.}~\bibnamefont {Parkin}},\ Vol.~\bibinfo {volume} {2}\
  (\bibinfo  {publisher} {John Wiley \& Sons},\ \bibinfo {address} {Chichester,
  UK},\ \bibinfo {year} {2007})\ p.\ \bibinfo {pages} {795–823}\BibitemShut
  {NoStop}%
\bibitem [{\citenamefont {Bessarab}\ \emph {et~al.}(2015)\citenamefont
  {Bessarab}, \citenamefont {Uzdin},\ and\ \citenamefont
  {Jónsson}}]{bessarab2015}%
  \BibitemOpen
  \bibfield  {author} {\bibinfo {author} {\bibfnamefont {P.~F.}\ \bibnamefont
  {Bessarab}}, \bibinfo {author} {\bibfnamefont {V.~M.}\ \bibnamefont
  {Uzdin}},\ and\ \bibinfo {author} {\bibfnamefont {H.}~\bibnamefont
  {Jónsson}},\ }\href
  {https://doi.org/https://doi.org/10.1016/j.cpc.2015.07.001} {\bibfield
  {journal} {\bibinfo  {journal} {Comput. Phys. Commun.}\ }\textbf {\bibinfo
  {volume} {196}},\ \bibinfo {pages} {335} (\bibinfo {year}
  {2015})}\BibitemShut {NoStop}%
\bibitem [{\citenamefont {Nocedal}\ and\ \citenamefont
  {Wright}(2006)}]{nocedal2006numerical}%
  \BibitemOpen
  \bibfield  {author} {\bibinfo {author} {\bibfnamefont {J.}~\bibnamefont
  {Nocedal}}\ and\ \bibinfo {author} {\bibfnamefont {S.}~\bibnamefont
  {Wright}},\ }\href@noop {} {\emph {\bibinfo {title} {Numerical
  optimization}}}\ (\bibinfo  {publisher} {Springer},\ \bibinfo {address} {New
  York},\ \bibinfo {year} {2006})\BibitemShut {NoStop}%
\bibitem [{\citenamefont {Ivanov}\ \emph {et~al.}(2021)\citenamefont {Ivanov},
  \citenamefont {Uzdin},\ and\ \citenamefont {J{\'o}nsson}}]{ivanov2021fast}%
  \BibitemOpen
  \bibfield  {author} {\bibinfo {author} {\bibfnamefont {A.}~\bibnamefont
  {Ivanov}}, \bibinfo {author} {\bibfnamefont {V.}~\bibnamefont {Uzdin}},\ and\
  \bibinfo {author} {\bibfnamefont {H.}~\bibnamefont {J{\'o}nsson}},\ }\href
  {https://doi.org/10.1016/j.cpc.2020.107749} {\bibfield  {journal} {\bibinfo
  {journal} {Comput. Phys. Commun.}\ }\textbf {\bibinfo {volume} {260}},\
  \bibinfo {pages} {107749} (\bibinfo {year} {2021})}\BibitemShut {NoStop}%
\bibitem [{\citenamefont {Deetz}\ and\ \citenamefont
  {Adams}(1945)}]{deetz1945elements}%
  \BibitemOpen
  \bibfield  {author} {\bibinfo {author} {\bibfnamefont {C.~H.}\ \bibnamefont
  {Deetz}}\ and\ \bibinfo {author} {\bibfnamefont {O.~S.}\ \bibnamefont
  {Adams}},\ }\href@noop {} {\emph {\bibinfo {title} {Elements of map
  projection}}}\ (\bibinfo  {publisher} {Citeseer},\ \bibinfo {year}
  {1945})\BibitemShut {NoStop}%
\bibitem [{\citenamefont {Mayer~Alegre}\ \emph {et~al.}(2007)\citenamefont
  {Mayer~Alegre}, \citenamefont {Torrezan},\ and\ \citenamefont
  {Medeiros-Ribeiro}}]{mayer2007microstrip}%
  \BibitemOpen
  \bibfield  {author} {\bibinfo {author} {\bibfnamefont {T.}~\bibnamefont
  {Mayer~Alegre}}, \bibinfo {author} {\bibfnamefont {A.}~\bibnamefont
  {Torrezan}},\ and\ \bibinfo {author} {\bibfnamefont {G.}~\bibnamefont
  {Medeiros-Ribeiro}},\ }\href {https://doi.org/10.1063/1.2809372} {\bibfield
  {journal} {\bibinfo  {journal} {Appl. Phys. Lett.}\ }\textbf {\bibinfo
  {volume} {91}},\ \bibinfo {pages} {204103} (\bibinfo {year}
  {2007})}\BibitemShut {NoStop}%
\bibitem [{\citenamefont {Curcic}\ \emph {et~al.}(2008)\citenamefont {Curcic},
  \citenamefont {Van~Waeyenberge}, \citenamefont {Vansteenkiste}, \citenamefont
  {Weigand}, \citenamefont {Sackmann}, \citenamefont {Stoll}, \citenamefont
  {F{\"a}hnle}, \citenamefont {Tyliszczak}, \citenamefont {Woltersdorf},
  \citenamefont {Back} \emph {et~al.}}]{curcic2008polarization}%
  \BibitemOpen
  \bibfield  {author} {\bibinfo {author} {\bibfnamefont {M.}~\bibnamefont
  {Curcic}}, \bibinfo {author} {\bibfnamefont {B.}~\bibnamefont
  {Van~Waeyenberge}}, \bibinfo {author} {\bibfnamefont {A.}~\bibnamefont
  {Vansteenkiste}}, \bibinfo {author} {\bibfnamefont {M.}~\bibnamefont
  {Weigand}}, \bibinfo {author} {\bibfnamefont {V.}~\bibnamefont {Sackmann}},
  \bibinfo {author} {\bibfnamefont {H.}~\bibnamefont {Stoll}}, \bibinfo
  {author} {\bibfnamefont {M.}~\bibnamefont {F{\"a}hnle}}, \bibinfo {author}
  {\bibfnamefont {T.}~\bibnamefont {Tyliszczak}}, \bibinfo {author}
  {\bibfnamefont {G.}~\bibnamefont {Woltersdorf}}, \bibinfo {author}
  {\bibfnamefont {C.~H.}\ \bibnamefont {Back}}, \emph {et~al.},\ }\href
  {https://doi.org/10.1103/PhysRevLett.101.197204} {\bibfield  {journal}
  {\bibinfo  {journal} {Phys. Rev. Lett}\ }\textbf {\bibinfo {volume} {101}},\
  \bibinfo {pages} {197204} (\bibinfo {year} {2008})}\BibitemShut {NoStop}%
\bibitem [{\citenamefont {Gao}\ \emph {et~al.}(2013)\citenamefont {Gao},
  \citenamefont {Lei}, \citenamefont {Chen}, \citenamefont {Xing},
  \citenamefont {Chen},\ and\ \citenamefont {Xie}}]{gao2013simple}%
  \BibitemOpen
  \bibfield  {author} {\bibinfo {author} {\bibfnamefont {H.}~\bibnamefont
  {Gao}}, \bibinfo {author} {\bibfnamefont {C.}~\bibnamefont {Lei}}, \bibinfo
  {author} {\bibfnamefont {M.}~\bibnamefont {Chen}}, \bibinfo {author}
  {\bibfnamefont {F.}~\bibnamefont {Xing}}, \bibinfo {author} {\bibfnamefont
  {H.}~\bibnamefont {Chen}},\ and\ \bibinfo {author} {\bibfnamefont
  {S.}~\bibnamefont {Xie}},\ }\href {https://doi.org/10.1364/OE.21.023107}
  {\bibfield  {journal} {\bibinfo  {journal} {Optics Express}\ }\textbf
  {\bibinfo {volume} {21}},\ \bibinfo {pages} {23107} (\bibinfo {year}
  {2013})}\BibitemShut {NoStop}%
\bibitem [{\citenamefont {Bisig}\ \emph {et~al.}(2015)\citenamefont {Bisig},
  \citenamefont {Mawass}, \citenamefont {St{\"a}rk}, \citenamefont {Moutafis},
  \citenamefont {Rhensius}, \citenamefont {Heidler}, \citenamefont {Gliga},
  \citenamefont {Weigand}, \citenamefont {Tyliszczak}, \citenamefont
  {Van~Waeyenberge} \emph {et~al.}}]{bisig2015dynamic}%
  \BibitemOpen
  \bibfield  {author} {\bibinfo {author} {\bibfnamefont {A.}~\bibnamefont
  {Bisig}}, \bibinfo {author} {\bibfnamefont {M.-A.}\ \bibnamefont {Mawass}},
  \bibinfo {author} {\bibfnamefont {M.}~\bibnamefont {St{\"a}rk}}, \bibinfo
  {author} {\bibfnamefont {C.}~\bibnamefont {Moutafis}}, \bibinfo {author}
  {\bibfnamefont {J.}~\bibnamefont {Rhensius}}, \bibinfo {author}
  {\bibfnamefont {J.}~\bibnamefont {Heidler}}, \bibinfo {author} {\bibfnamefont
  {S.}~\bibnamefont {Gliga}}, \bibinfo {author} {\bibfnamefont
  {M.}~\bibnamefont {Weigand}}, \bibinfo {author} {\bibfnamefont
  {T.}~\bibnamefont {Tyliszczak}}, \bibinfo {author} {\bibfnamefont
  {B.}~\bibnamefont {Van~Waeyenberge}}, \emph {et~al.},\ }\href
  {https://doi.org/10.1063/1.4915256} {\bibfield  {journal} {\bibinfo
  {journal} {Appl. Phys. Lett.}\ }\textbf {\bibinfo {volume} {106}},\ \bibinfo
  {pages} {122401} (\bibinfo {year} {2015})}\BibitemShut {NoStop}%
\bibitem [{\citenamefont {Rius}\ \emph {et~al.}(2016)\citenamefont {Rius},
  \citenamefont {Bolea}, \citenamefont {Mora},\ and\ \citenamefont
  {Capmany}}]{rius2016incoherent}%
  \BibitemOpen
  \bibfield  {author} {\bibinfo {author} {\bibfnamefont {M.}~\bibnamefont
  {Rius}}, \bibinfo {author} {\bibfnamefont {M.}~\bibnamefont {Bolea}},
  \bibinfo {author} {\bibfnamefont {J.}~\bibnamefont {Mora}},\ and\ \bibinfo
  {author} {\bibfnamefont {J.}~\bibnamefont {Capmany}},\ }\href
  {https://doi.org/10.1109/LPT.2016.2623360} {\bibfield  {journal} {\bibinfo
  {journal} {IEEE Photonics Technology Letters}\ }\textbf {\bibinfo {volume}
  {29}},\ \bibinfo {pages} {7} (\bibinfo {year} {2016})}\BibitemShut {NoStop}%
\bibitem [{\citenamefont {Vlasov}\ \emph {et~al.}(2022)\citenamefont {Vlasov},
  \citenamefont {Kwiatkowski}, \citenamefont {Lobanov}, \citenamefont {Uzdin},\
  and\ \citenamefont {Bessarab}}]{Vlasov2022}%
  \BibitemOpen
  \bibfield  {author} {\bibinfo {author} {\bibfnamefont {S.~M.}\ \bibnamefont
  {Vlasov}}, \bibinfo {author} {\bibfnamefont {G.~J.}\ \bibnamefont
  {Kwiatkowski}}, \bibinfo {author} {\bibfnamefont {I.~S.}\ \bibnamefont
  {Lobanov}}, \bibinfo {author} {\bibfnamefont {V.~M.}\ \bibnamefont {Uzdin}},\
  and\ \bibinfo {author} {\bibfnamefont {P.~F.}\ \bibnamefont {Bessarab}},\
  }\href {https://doi.org/10.1103/PhysRevB.105.134404} {\bibfield  {journal}
  {\bibinfo  {journal} {Phys. Rev. B}\ }\textbf {\bibinfo {volume} {105}},\
  \bibinfo {pages} {134404} (\bibinfo {year} {2022})}\BibitemShut {NoStop}%
\bibitem [{\citenamefont {Bessarab}\ \emph {et~al.}(2012)\citenamefont
  {Bessarab}, \citenamefont {Uzdin},\ and\ \citenamefont
  {J\'onsson}}]{bessarab2012harmonic}%
  \BibitemOpen
  \bibfield  {author} {\bibinfo {author} {\bibfnamefont {P.~F.}\ \bibnamefont
  {Bessarab}}, \bibinfo {author} {\bibfnamefont {V.~M.}\ \bibnamefont
  {Uzdin}},\ and\ \bibinfo {author} {\bibfnamefont {H.}~\bibnamefont
  {J\'onsson}},\ }\href {https://doi.org/10.1103/PhysRevB.85.184409} {\bibfield
   {journal} {\bibinfo  {journal} {Phys. Rev. B}\ }\textbf {\bibinfo {volume}
  {85}},\ \bibinfo {pages} {184409} (\bibinfo {year} {2012})}\BibitemShut
  {NoStop}%
\bibitem [{\citenamefont {Fiedler}\ \emph {et~al.}(2012)\citenamefont
  {Fiedler}, \citenamefont {Fidler}, \citenamefont {Lee}, \citenamefont
  {Schrefl}, \citenamefont {Stamps}, \citenamefont {Braun},\ and\ \citenamefont
  {Suess}}]{Fiedler2012}%
  \BibitemOpen
  \bibfield  {author} {\bibinfo {author} {\bibfnamefont {G.}~\bibnamefont
  {Fiedler}}, \bibinfo {author} {\bibfnamefont {J.}~\bibnamefont {Fidler}},
  \bibinfo {author} {\bibfnamefont {J.}~\bibnamefont {Lee}}, \bibinfo {author}
  {\bibfnamefont {T.}~\bibnamefont {Schrefl}}, \bibinfo {author} {\bibfnamefont
  {R.~L.}\ \bibnamefont {Stamps}}, \bibinfo {author} {\bibfnamefont {H.~B.}\
  \bibnamefont {Braun}},\ and\ \bibinfo {author} {\bibfnamefont
  {D.}~\bibnamefont {Suess}},\ }\href {https://doi.org/10.1063/1.4712033}
  {\bibfield  {journal} {\bibinfo  {journal} {J. of Appl. Phys.}\ }\textbf
  {\bibinfo {volume} {111}},\ \bibinfo {pages} {093917} (\bibinfo {year}
  {2012})}\BibitemShut {NoStop}%
\bibitem [{\citenamefont {Braun}(1993)}]{Braun1993}%
  \BibitemOpen
  \bibfield  {author} {\bibinfo {author} {\bibfnamefont {H.-B.}\ \bibnamefont
  {Braun}},\ }\href {https://doi.org/10.1103/PhysRevLett.71.3557} {\bibfield
  {journal} {\bibinfo  {journal} {Phys. Rev. Lett.}\ }\textbf {\bibinfo
  {volume} {71}},\ \bibinfo {pages} {3557} (\bibinfo {year}
  {1993})}\BibitemShut {NoStop}%
\bibitem [{\citenamefont {Braun}(1994)}]{Braun1994}%
  \BibitemOpen
  \bibfield  {author} {\bibinfo {author} {\bibfnamefont {H.-B.}\ \bibnamefont
  {Braun}},\ }\href {https://doi.org/10.1103/PhysRevB.50.16501} {\bibfield
  {journal} {\bibinfo  {journal} {Phys. Rev. B}\ }\textbf {\bibinfo {volume}
  {50}},\ \bibinfo {pages} {16501} (\bibinfo {year} {1994})}\BibitemShut
  {NoStop}%
\bibitem [{\citenamefont {Bode}\ \emph {et~al.}(2004)\citenamefont {Bode},
  \citenamefont {Pietzsch}, \citenamefont {Kubetzka},\ and\ \citenamefont
  {Wiesendanger}}]{bode2004shape}%
  \BibitemOpen
  \bibfield  {author} {\bibinfo {author} {\bibfnamefont {M.}~\bibnamefont
  {Bode}}, \bibinfo {author} {\bibfnamefont {O.}~\bibnamefont {Pietzsch}},
  \bibinfo {author} {\bibfnamefont {A.}~\bibnamefont {Kubetzka}},\ and\
  \bibinfo {author} {\bibfnamefont {R.}~\bibnamefont {Wiesendanger}},\ }\href
  {https://doi.org/10.1103/PhysRevLett.92.067201} {\bibfield  {journal}
  {\bibinfo  {journal} {Phys. Rev. Lett.}\ }\textbf {\bibinfo {volume} {92}},\
  \bibinfo {pages} {067201} (\bibinfo {year} {2004})}\BibitemShut {NoStop}%
\bibitem [{\citenamefont {Krause}\ \emph {et~al.}(2009)\citenamefont {Krause},
  \citenamefont {Herzog}, \citenamefont {Stapelfeldt}, \citenamefont
  {Berbil-Bautista}, \citenamefont {Bode}, \citenamefont {Vedmedenko},\ and\
  \citenamefont {Wiesendanger}}]{krause2009magnetization}%
  \BibitemOpen
  \bibfield  {author} {\bibinfo {author} {\bibfnamefont {S.}~\bibnamefont
  {Krause}}, \bibinfo {author} {\bibfnamefont {G.}~\bibnamefont {Herzog}},
  \bibinfo {author} {\bibfnamefont {T.}~\bibnamefont {Stapelfeldt}}, \bibinfo
  {author} {\bibfnamefont {L.}~\bibnamefont {Berbil-Bautista}}, \bibinfo
  {author} {\bibfnamefont {M.}~\bibnamefont {Bode}}, \bibinfo {author}
  {\bibfnamefont {E.~Y.}\ \bibnamefont {Vedmedenko}},\ and\ \bibinfo {author}
  {\bibfnamefont {R.}~\bibnamefont {Wiesendanger}},\ }\href
  {https://doi.org/10.1103/PhysRevLett.103.127202} {\bibfield  {journal}
  {\bibinfo  {journal} {Phys. Rev. Lett.}\ }\textbf {\bibinfo {volume} {103}},\
  \bibinfo {pages} {127202} (\bibinfo {year} {2009})}\BibitemShut {NoStop}%
\bibitem [{\citenamefont {Bessarab}\ \emph {et~al.}(2013)\citenamefont
  {Bessarab}, \citenamefont {Uzdin},\ and\ \citenamefont
  {J\'onsson}}]{bessarab2013size}%
  \BibitemOpen
  \bibfield  {author} {\bibinfo {author} {\bibfnamefont {P.~F.}\ \bibnamefont
  {Bessarab}}, \bibinfo {author} {\bibfnamefont {V.~M.}\ \bibnamefont
  {Uzdin}},\ and\ \bibinfo {author} {\bibfnamefont {H.}~\bibnamefont
  {J\'onsson}},\ }\href {https://doi.org/10.1103/PhysRevLett.110.020604}
  {\bibfield  {journal} {\bibinfo  {journal} {Phys. Rev. Lett.}\ }\textbf
  {\bibinfo {volume} {110}},\ \bibinfo {pages} {020604} (\bibinfo {year}
  {2013})}\BibitemShut {NoStop}%
\bibitem [{\citenamefont {Seki}\ \emph {et~al.}(2013)\citenamefont {Seki},
  \citenamefont {Utsumiya}, \citenamefont {Nozaki}, \citenamefont {Imamura},\
  and\ \citenamefont {Takanashi}}]{seki2013spin}%
  \BibitemOpen
  \bibfield  {author} {\bibinfo {author} {\bibfnamefont {T.}~\bibnamefont
  {Seki}}, \bibinfo {author} {\bibfnamefont {K.}~\bibnamefont {Utsumiya}},
  \bibinfo {author} {\bibfnamefont {Y.}~\bibnamefont {Nozaki}}, \bibinfo
  {author} {\bibfnamefont {H.}~\bibnamefont {Imamura}},\ and\ \bibinfo {author}
  {\bibfnamefont {K.}~\bibnamefont {Takanashi}},\ }\href
  {https://doi.org/10.1038/ncomms2737} {\bibfield  {journal} {\bibinfo
  {journal} {Nat. Commun.}\ }\textbf {\bibinfo {volume} {4}},\ \bibinfo {pages}
  {1726} (\bibinfo {year} {2013})}\BibitemShut {NoStop}%
\bibitem [{\citenamefont {Spinelli}\ \emph {et~al.}(2014)\citenamefont
  {Spinelli}, \citenamefont {Bryant}, \citenamefont {Delgado}, \citenamefont
  {Fern{\'a}ndez-Rossier},\ and\ \citenamefont {Otte}}]{Spinelli2014}%
  \BibitemOpen
  \bibfield  {author} {\bibinfo {author} {\bibfnamefont {A.}~\bibnamefont
  {Spinelli}}, \bibinfo {author} {\bibfnamefont {B.}~\bibnamefont {Bryant}},
  \bibinfo {author} {\bibfnamefont {F.}~\bibnamefont {Delgado}}, \bibinfo
  {author} {\bibfnamefont {J.}~\bibnamefont {Fern{\'a}ndez-Rossier}},\ and\
  \bibinfo {author} {\bibfnamefont {A.~F.}\ \bibnamefont {Otte}},\ }\href
  {https://doi.org/10.1038/nmat4018} {\bibfield  {journal} {\bibinfo  {journal}
  {Nat. Mater.}\ }\textbf {\bibinfo {volume} {13}},\ \bibinfo {pages} {782}
  (\bibinfo {year} {2014})}\BibitemShut {NoStop}%
\bibitem [{\citenamefont {Yamaji}\ and\ \citenamefont
  {Imamura}(2018)}]{Yamaji2018}%
  \BibitemOpen
  \bibfield  {author} {\bibinfo {author} {\bibfnamefont {T.}~\bibnamefont
  {Yamaji}}\ and\ \bibinfo {author} {\bibfnamefont {H.}~\bibnamefont
  {Imamura}},\ }\href {https://doi.org/10.1063/1.5029219} {\bibfield  {journal}
  {\bibinfo  {journal} {Appl. Phys. Lett.}\ }\textbf {\bibinfo {volume}
  {112}},\ \bibinfo {pages} {202406} (\bibinfo {year} {2018})}\BibitemShut
  {NoStop}%
\bibitem [{\citenamefont {Mentink}\ \emph {et~al.}(2010)\citenamefont
  {Mentink}, \citenamefont {Tretyakov}, \citenamefont {Fasolino}, \citenamefont
  {Katsnelson},\ and\ \citenamefont {Rasing}}]{mentink2010stable}%
  \BibitemOpen
  \bibfield  {author} {\bibinfo {author} {\bibfnamefont {J.}~\bibnamefont
  {Mentink}}, \bibinfo {author} {\bibfnamefont {M.}~\bibnamefont {Tretyakov}},
  \bibinfo {author} {\bibfnamefont {A.}~\bibnamefont {Fasolino}}, \bibinfo
  {author} {\bibfnamefont {M.}~\bibnamefont {Katsnelson}},\ and\ \bibinfo
  {author} {\bibfnamefont {T.}~\bibnamefont {Rasing}},\ }\href
  {https://doi.org/10.1088/0953-8984/22/17/176001} {\bibfield  {journal}
  {\bibinfo  {journal} {J. Phys. Condens. Matter}\ }\textbf {\bibinfo {volume}
  {22}},\ \bibinfo {pages} {176001} (\bibinfo {year} {2010})}\BibitemShut
  {NoStop}%
\bibitem [{\citenamefont {Richter}(2009)}]{richter2009density}%
  \BibitemOpen
  \bibfield  {author} {\bibinfo {author} {\bibfnamefont {H.}~\bibnamefont
  {Richter}},\ }\href {https://doi.org/10.1016/j.jmmm.2008.04.161} {\bibfield
  {journal} {\bibinfo  {journal} {J. Magn. Magn. Mater}\ }\textbf {\bibinfo
  {volume} {321}},\ \bibinfo {pages} {467} (\bibinfo {year}
  {2009})}\BibitemShut {NoStop}%
\bibitem [{\citenamefont {Krounbi}\ \emph {et~al.}(2015)\citenamefont
  {Krounbi}, \citenamefont {Nikitin}, \citenamefont {Apalkov}, \citenamefont
  {Lee}, \citenamefont {Tang}, \citenamefont {Beach}, \citenamefont
  {Erickson},\ and\ \citenamefont {Chen}}]{krounbi2015keynote}%
  \BibitemOpen
  \bibfield  {author} {\bibinfo {author} {\bibfnamefont {M.}~\bibnamefont
  {Krounbi}}, \bibinfo {author} {\bibfnamefont {V.}~\bibnamefont {Nikitin}},
  \bibinfo {author} {\bibfnamefont {D.}~\bibnamefont {Apalkov}}, \bibinfo
  {author} {\bibfnamefont {J.}~\bibnamefont {Lee}}, \bibinfo {author}
  {\bibfnamefont {X.}~\bibnamefont {Tang}}, \bibinfo {author} {\bibfnamefont
  {R.}~\bibnamefont {Beach}}, \bibinfo {author} {\bibfnamefont
  {D.}~\bibnamefont {Erickson}},\ and\ \bibinfo {author} {\bibfnamefont
  {E.}~\bibnamefont {Chen}},\ }\href {https://doi.org/10.1149/06903.0119ecst}
  {\bibfield  {journal} {\bibinfo  {journal} {ECS Transactions}\ }\textbf
  {\bibinfo {volume} {69}},\ \bibinfo {pages} {119} (\bibinfo {year}
  {2015})}\BibitemShut {NoStop}%
\end{thebibliography}%

\end{document}